\documentclass[twocolumn]{emulateapj} 
\pdfoutput=1
\usepackage{subfigure,graphicx}
\usepackage{url}
\usepackage{color}
\definecolor{patinared}{rgb}{.72,0,0}
\definecolor{patinablue}{rgb}{0,0,.65}
\definecolor{forest}{RGB}{8,120,4} 
\usepackage[bookmarks=false]{hyperref}
\hypersetup{
  colorlinks= true, 
  urlcolor  = black, 
  linkcolor = patinared, 
  citecolor = patinablue 
}
\usepackage{epstopdf}

\newcommand{\lt}{\textsc{Little Things}}
\newcommand{\herschel}{\textit {Herschel}}
\newcommand{\spitzer}{\textit {Spitzer}}
\newcommand{\iras}{IRAS} 
\newcommand{\iso}{ISO} 
\newcommand{\kao}{\textit {KAO}}
\newcommand{\cii}{[\ion{C}{2}]}
\newcommand{\oi}{[\ion{O}{1}]}
\newcommand{\oiii}{[\ion{O}{3}]}
\newcommand{\nii}{[\ion{N}{2}]}
\newcommand{\hi}{\ion{H}{1}}
\newcommand{\hii}{\ion{H}{2}}

\newcommand{\spc}{\texttt{slicedProjectedCubes}}

\definecolor{orange}{rgb}{1,0.3,0}

\usepackage[normalem]{ulem}

\defcitealias{Brauher}{B+08}
\defcitealias{HunterISO}{H+01}
\defcitealias{HunterHa}{HE04}
\defcitealias{HunterUBV}{HE06}
\defcitealias{Mal01}{M+01}
\defcitealias{Cormier2015}{C+15}

\slugcomment{Accepted for publication in The Astronomical Journal }
\shorttitle{\herschel\ Spectroscopy of \lt\ Dwarfs}   
\shortauthors{Cigan et al.} 

\begin{document}

\title{Herschel Spectroscopic Observations of \lt\ Dwarf Galaxies}

\author{Phil Cigan\altaffilmark{1}, Lisa Young\altaffilmark{1}, Diane Cormier\altaffilmark{2}, Vianney Lebouteiller\altaffilmark{3}, Suzanne Madden\altaffilmark{3}, Deidre Hunter\altaffilmark{4}, Elias Brinks\altaffilmark{5}, Bruce Elmegreen\altaffilmark{6}, Andreas Schruba\altaffilmark{7,8}, Volker Heesen\altaffilmark{9}, and the \lt\ Team}
\altaffiltext{1}{Physics Department, New Mexico Institute of Mining and Technology, Socorro, NM 87801, USA; pcigan@nmt.edu}
\altaffiltext{2}{Institut f\"ur Theoretische Astrophysik, Zentrum f\"ur Astronomie der Universit\"at Heidelberg, Albert-Ueberle Str. 2, D-69120 Heidelberg, Germany}
\altaffiltext{3}{Laboratoire AIM, CEA/DSM - CNRS - Universit\'e ́Paris Diderot, Irfu/Service d'Astrophysique, CEA Saclay, 91191 Gif-sur-Yvette, France}
\altaffiltext{4}{Lowell Observatory, 1400 West Mars Hill Road, Flagstaff, AZ 86001, USA}
\altaffiltext{5}{Centre for Astrophysics Research, University of Hertfordshire, College Lane, Hatfield, AL10 9AB, UK}
\altaffiltext{6}{IBM T.J. Watson Research Center, 1101 Kitchawan Road, Yorktown Hts., NY 10598, USA }
\altaffiltext{7}{Max-Planck-Institut f\"ur extraterrestrische Physik, Giessenbachstrasse 1, 85748 Garching, Germany}
\altaffiltext{8}{Center for Astronomy and Astrophysics, California Institute of Technology, 1200 E California Blvd, MC 249-17, Pasadena, CA 91125, USA }
\altaffiltext{9}{School of Physics and Astronomy, University of Southampton, Southampton SO17 1BJ, UK}

%
%
%
%
%

\begin{abstract}
We present far-infrared spectral line observations of five galaxies from the \lt\ sample: DDO 69, DDO 70, DDO 75, DDO 155, and WLM.  While most studies of dwarfs focus on bright systems or starbursts due to observational constraints, our data extend the observed parameter space into the regime of low surface brightness dwarf galaxies with low metallicities and moderate star formation rates.  Our targets were observed with \herschel\ at the \cii\ 158$\mu$m, \oi\ 63$\mu$m, \oiii\ 88$\mu$m, and \nii\ 122$\mu$m emission lines using the PACS Spectrometer.  These high-resolution maps allow us for the first time to study the far-infrared properties of these systems on the scales of larger star-forming complexes. The spatial resolution in our maps, in combination with star formation tracers, allows us to identify separate PDRs in some of the regions we observed. Our systems have widespread \cii\ emission that is bright relative to continuum, averaging near 0.5\% of the total infrared budget - higher than in solar-metallicity galaxies of other types.  \nii\ is weak, suggesting that the \cii\ emission in our galaxies comes mostly from PDRs instead of the diffuse ionized ISM.  These systems exhibit efficient cooling at low dust temperatures, as shown by (\oi+\cii)/TIR in relation to 60$\mu$m/100$\mu$m, and low \oi/\cii\ ratios which indicate that \cii\ is the dominant coolant of the ISM.  We observe \oiii/\cii\ ratios in our galaxies that are lower than those published for other dwarfs, but similar to levels noted in spirals.

\end{abstract}


\section{INTRODUCTION}

Understanding how conditions in the interstellar medium (ISM) influence star formation is crucial for models and simulations of galaxy evolution.  To address the issue properly, we must understand a wide variety of environments, from mergers to spirals to dwarf galaxies.  Dwarfs in particular are interesting because of their typically low abundances of metals, which can lead to quite different physical conditions in the ISM of these systems. Molecular gas, H$_2$, is the fuel necessary for star formation, but is easily dissociated by UV radiation.  In metal-rich environments, the relatively high abundance of dust can better shield H$_2$.  Thus observations of the stuctures of molecular clouds and their envelopes in dwarf galaxies can shed light on the process of star formation in the low-metallicity regime.  Since molecular gas is difficult to study directly in low-metallicity systems, we utilize far-infrared (FIR) atomic spectral lines to determine the properties of the ISM and photodissociation regions (PDRs).

The FIR fine-structure lines arising from star-formation regions (\hii\ regions and PDRs) are some of the most important coolants in these galaxies, and can reveal the physical conditions and structure of their ISM.  For example, the \cii\ 158$\mu$m line can range from 0.1\% to 1\% of the FIR luminosity in galaxies (e.g., \citealt{Stacey91,Brauher}).  Other important cooling lines include \oi\ 63$\mu$m and 145$\mu$m, for example.  Since each atomic species exists within a certain range of densities and levels of ionizing radiation, measured abundances and ratios of these diagnostic lines can be used to determine the local gas properties such as density and local radiation field.  

Starting with the \textit{Kuiper} Airborne Observatory (\kao), the \cii\ line at 158$\mu$m was found to be one of the brightest FIR spectral lines in external galaxies \citep[e.g.,][]{Crawford85,Stacey91}, though the sample was small, the accessible lines were few, and sensitivity was poor.  The Infrared Space Observatory (\iso) allowed for larger and improved surveys of FIR lines in galaxies \citep[see, for example,][]{Mal97,SmithMadden97,Brauher}. When dwarfs were observed with \kao\ and \iso\ \citep[see][for further discussion]{Madden2000,HunterISO}, it was found that they tend to have higher \cii/FIR ratios than spirals. However, due to sensitivity limitations, large numbers of dwarf galaxies were not accessible at FIR wavelengths. 

Now, with the enhanced spatial resolution of the \textit{Herschel} Space Observatory \citep{Pil10}, we can expand on previous studies that were limited to average fluxes over entire galaxies.  The sensitivity of the PACS spectrometer \citep{Pog10} allows us to actually survey the FIR lines in lower-metallicity systems, and its high spatial resolution of 11.5\arcsec\ \citep{Aniano11} allows us to study separate resolved regions within each image.  Other spectral studies of low-metallicity systems with \herschel\ include the Dwarf Galaxy Survey \citep[DGS;][]{DGS}, which targeted 50 metal-poor dwarfs as low as 2\% solar metallicity -- among the lowest-metallicity galaxies in the local universe.  Detailed models of PDRs in these systems have been developed for NGC4214 \citep{Dimaratos2015,Cor10}, Haro 11 \citep{Cor12} and the N11 region of the LMC \citep{Leb12}.  While previously studied systems at low metallicities in the DGS and other samples include objects with moderate star formation rates, many form stars at a high rate. An extreme example is Haro 11, which is considered to be in a starburst phase with a SFR of 43 M$_\odot$ yr$^{-1}$ \citep{DeLooze2014}. 
We expand the sample of observed systems to include more galaxies representative of typical dwarfs.

To determine the structure of star-forming molecular clouds at low metallicity and moderate star formation rates, we have mapped far-infrared (FIR) fine-structure lines in selected regions of five dwarf irregular galaxies: DDO 69, DDO 70, DDO 75, DDO 155, and WLM.  This sample is part of the larger \lt\ \citep[Local Irregulars That Trace Luminosity Extremes, The HI Nearby Galaxy Survey;][]{LTdata} sample of 41 nearby dwarf galaxies.  \lt\ is a survey of relatively normal, nearby gas-rich dwarf galaxies, with an impressive suite of data amassed to trace their stellar populations, gas content, dynamics, and star formation indicators. The \lt\ sample was itself selected from a larger sample of 136 dwarfs discussed by \citet[][hereafter \citetalias{HunterHa} and \citetalias{HunterUBV}]{HunterHa,HunterUBV}, and distributions of parameters such as gas mass, central surface brightness, and star formation rate for both \lt\ and its parent sample are presented in Figure 1 of \cite{LTdata}.  Our specific targets were selected to be representative of a  large number of dwarfs in terms of their SFR and metallicity. The average star formation rate (SFR$_{FUV}$) in our \herschel\ sample is a modest 0.095 M$_\sun$ yr$^{-1}$, and the average gas depletion time (M$_{\mathrm{HI}}$/SFR) is roughly 8Gyr. The sample metallicities span from $\sim$ 13\% Z$_\sun$ down to $\sim$ 5\% Z$_\sun$ (where Z$_\sun$ defined by 12 + log($\mathrm{O}/\mathrm{H}$) is 8.69, \citealp{Asplund09}), complementing the sample observed by DGS, which has FIR observations of galaxies as low as Z=7.14, or 3\% solar metallicity.  Our targets, all less than 2.2 Mpc away, are extended, low surface brightness galaxies with only moderate star formation.  Their close proximity allows us to study them on scales down to 45-123 pc (corresponding to the 11\farcs5 \cii\ angular resolution at distances of 0.8-2.2 Mpc).  Basic information about each galaxy in our sample is presented in Table ~\ref{table:basicinfo}. 

In this work we quantify the emission recovered from several infrared spectral lines -- \cii\ 157.74$\mu$m, \oi\ 63.18$\mu$m, \nii\ 121.90$\mu$m, and \oiii\ 88.36$\mu$m -- and their flux ratios.  We also compare the line fluxes with integrated total infrared (TIR) emission.  We discuss our results in the context of previous work in the literature and comment on the physical conditions determined from these observations in our target regions.  We present two different viewpoints on the regions in our galaxies: we investigate flux values integrated over the whole field of view, and we analyze our maps on a resolved pixel-by-pixel basis.  The results of this work give essential input for modeling the PDRs in these systems, which we will explore in an upcoming paper.

\begin{figure}
\label{fig:zoom}
\includegraphics[width=3.5in]{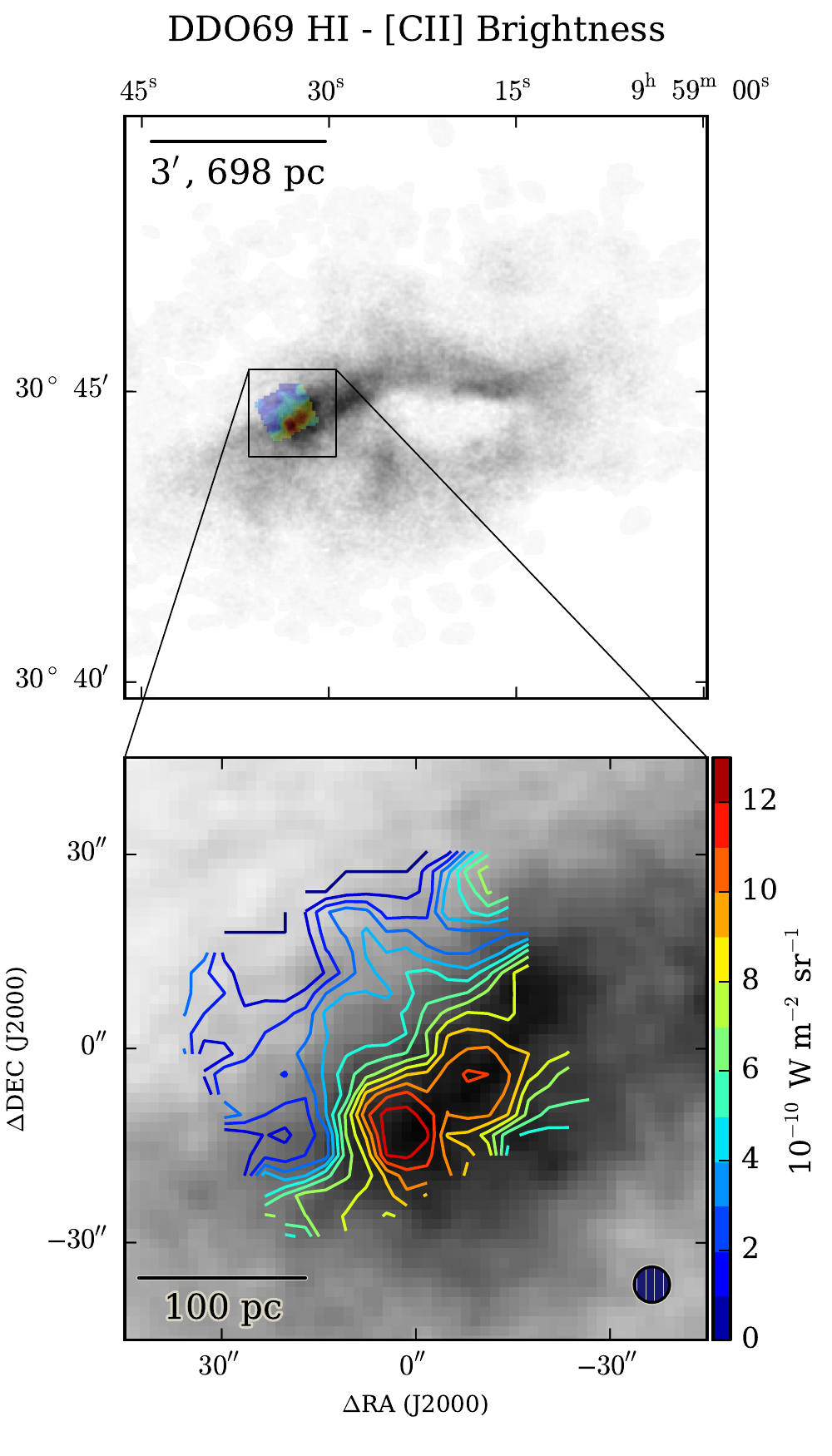}
\caption{Overlay of \herschel\ \cii\ contours onto \hi\ in DDO 69. Darker gray corresponds to increased \hi\ emission.  The \cii\ beam is shown in the lower frame.  Overlays of the other galaxies can be found in Figure~\ref{fig:zoomonline}.  }
\end{figure}

\begin{figure*}
\centering
	\begin{subfigure}[]{}
		\includegraphics[width=0.35\textwidth]{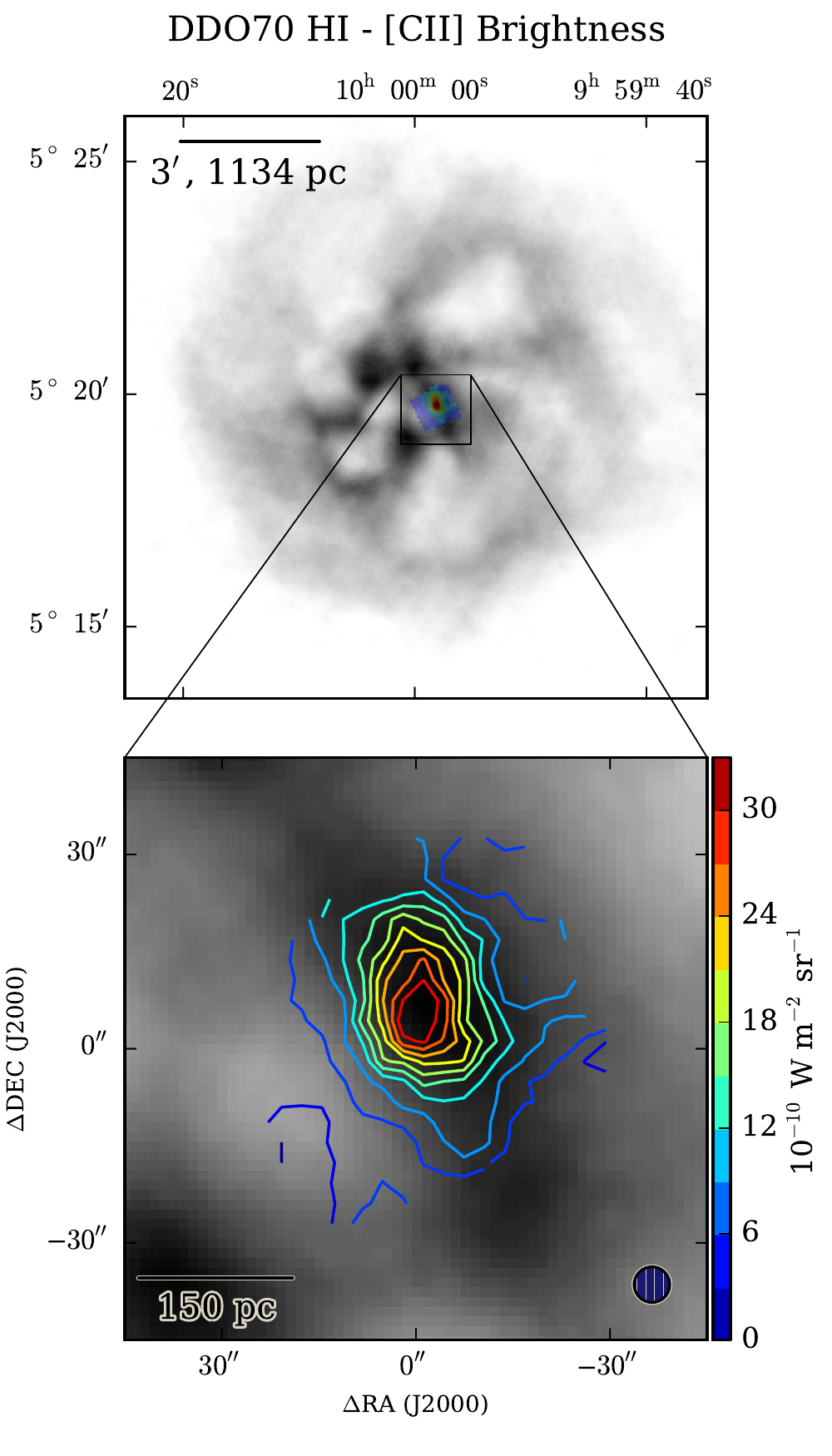}
	\end{subfigure}	
	\begin{subfigure}[]{}
		\includegraphics[width=0.35\textwidth]{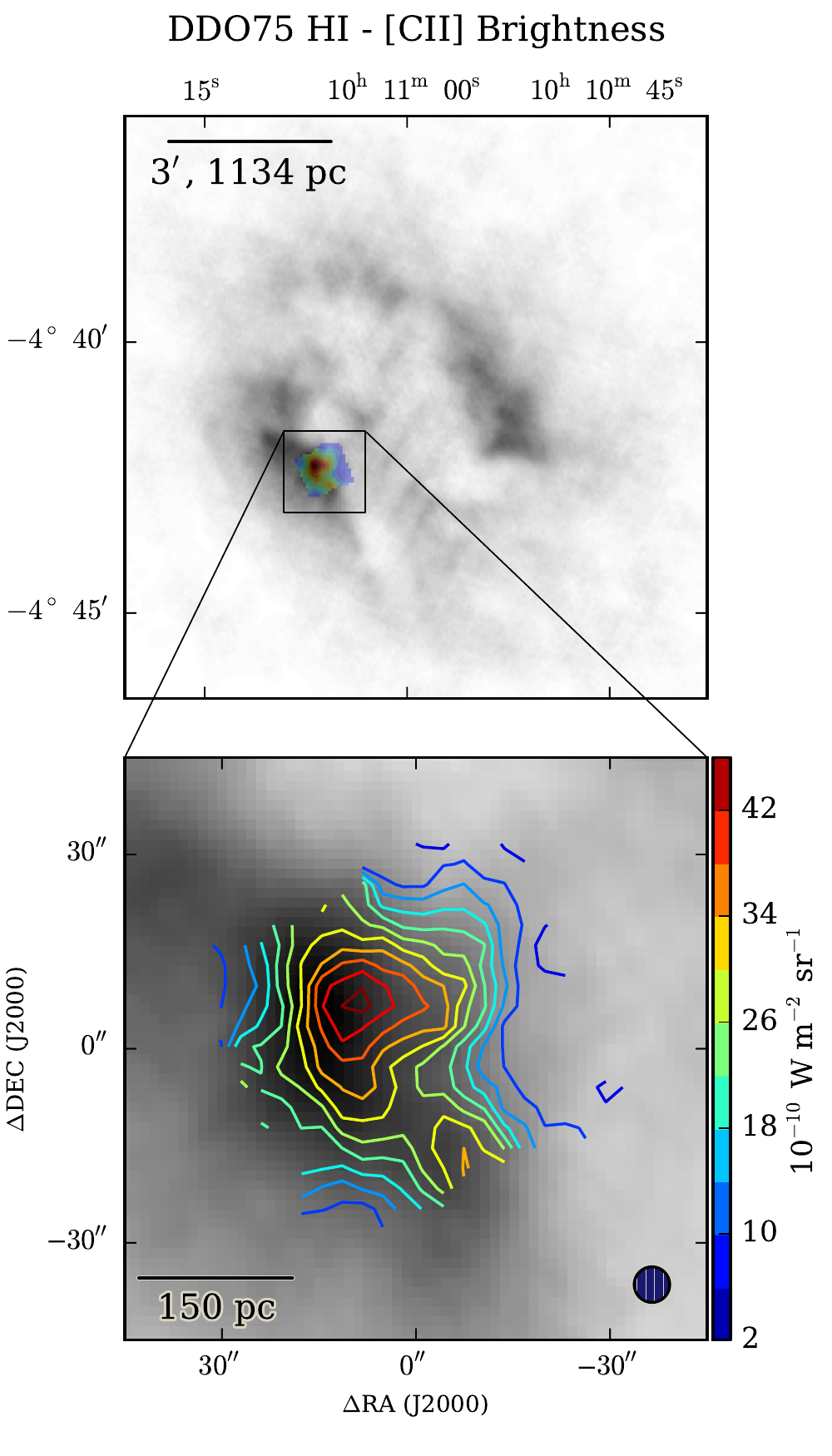}
	\end{subfigure}	
	\\
	\begin{subfigure}[]{}
		\includegraphics[width=0.35\textwidth]{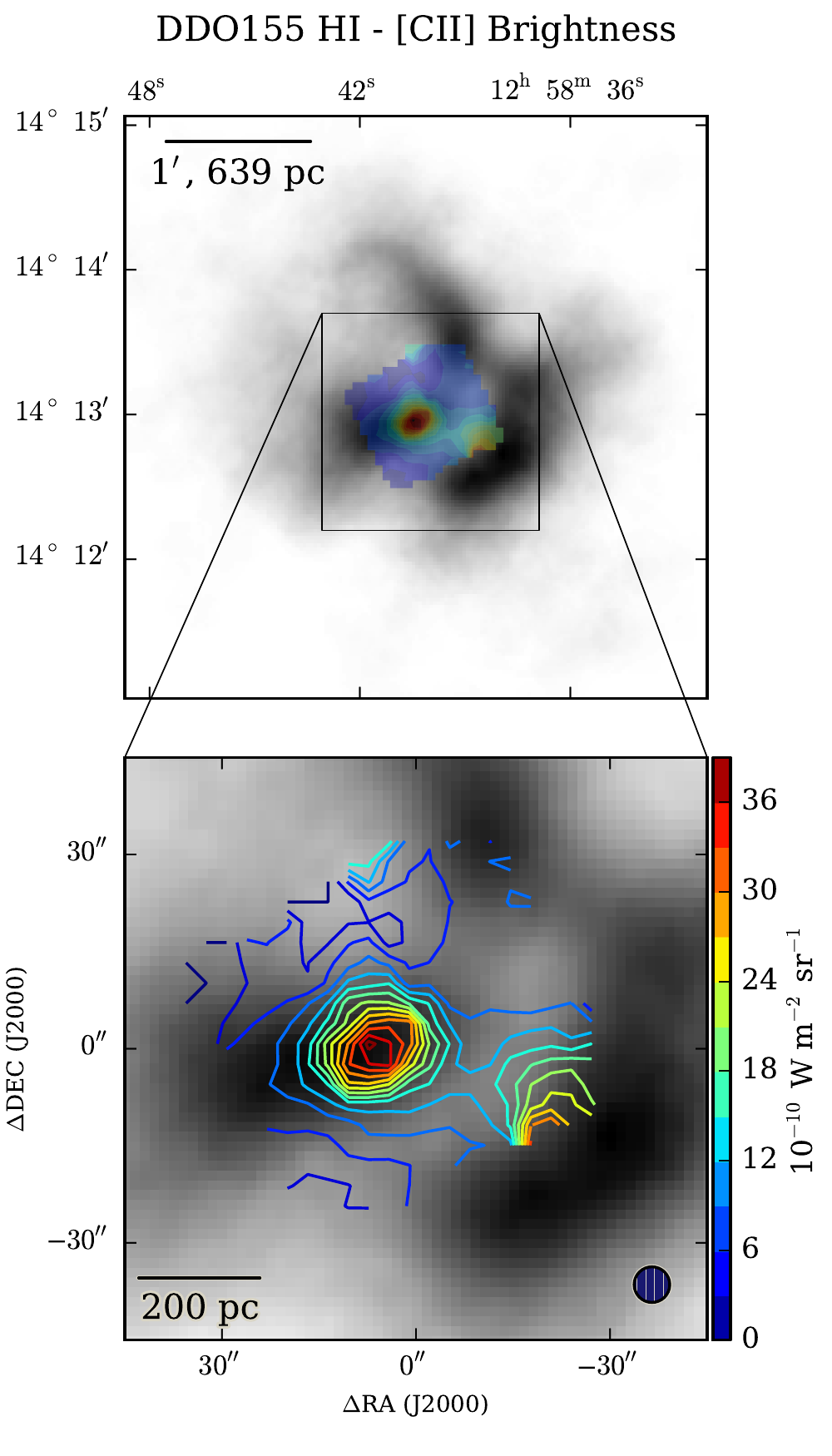}
	\end{subfigure}	
	\begin{subfigure}[]{}
		\includegraphics[width=0.35\textwidth]{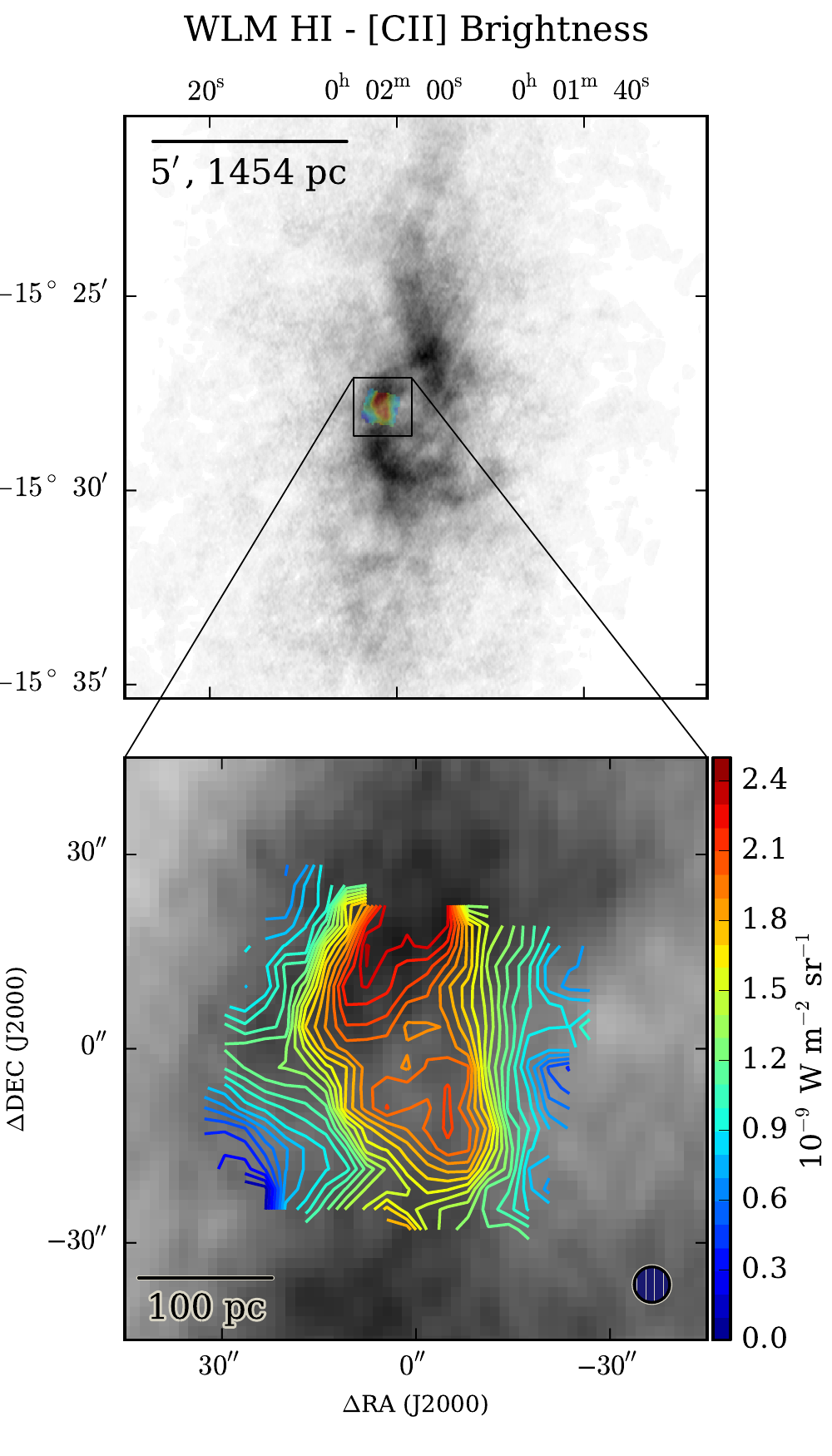}
	\end{subfigure}
\caption{ Overlays of \herschel\ \cii\ contours onto \hi\ maps from \cite{LTdata}. Each PACS footprint covers 51\farcs 5 on a side.   Darkness in the grayscale corresponds to increased \hi\ emission. The contour levels and linear scales are different for each target.}
\label{fig:zoomonline}
\end{figure*}

\begin{deluxetable*}{ llccccccc }
\tabletypesize{\footnotesize}

\tablecaption{ Basic Galaxy Information \label{table:basicinfo}} 

\tablehead{ \colhead{Galaxy} & \colhead{Other Names} & \colhead{D} & \colhead{log$_{10}$ M$_\mathrm{HI}$} & \colhead{R$_D$} & \colhead{${\mu_0}^V$} & \colhead{log$_{10}$ SFR$^{FUV}$} & \colhead{log$_{10}$ SFR$_D^{FUV}$} & \colhead{12+log$_{10}$(O/H)} \\  
 & & (Mpc) & (M$_\odot$) & (kpc) & (mag arcsec$^{-2}$) & (M$_\sun \; \textrm{yr}^{-1}$) & (M$_\sun \; \textrm{yr}^{-1} \; \textrm{kpc}^{-2}$) & }

\startdata
DDO 69 & PGC 28868 & 0.8 & 6.84 & 0.19 $\pm$ 0.01 & 23.01 & -3.17 & -2.22 $\pm$ 0.01 & 7.38 $\pm$ 0.10 \\
       & UGC 5364 \\
       & Leo A    \\
DDO 70 & PGC 28913 & 1.3 & 7.61 & 0.48 $\pm$ 0.01 & 23.81 & -2.30 & -2.16 $\pm$ 0.00 & 7.53 $\pm$ 0.06 \\
       & UGC 5373 \\
       & Sextans B\\
DDO 75 & PGC 29653 & 1.3 & 7.86 & 0.22 $\pm$ 0.01 & 20.40 & -1.89 & -1.07 $\pm$ 0.01 & 7.54 $\pm$ 0.06 \\
       & UGCA 205 \\
       & Sextans A\\
DDO 155 & PGC 44491 & 2.2 & 7.00 & 0.15 $\pm$ 0.01 & 21.72 & \nodata & \nodata & 7.65 $\pm$ 0.06 \\
        & UGC 8091  \\
        & GR 8      \\
        & VII Zw 222\\
WLM & PGC 143 & 1.0 & 7.85 & 0.57 $\pm$ 0.03 & 22.97 & -2.04 & -2.05 $\pm$ 0.01 & 7.83 $\pm$ 0.06 \\
    & UGCA 444             \\
    & DDO 221              

\\ \multicolumn{9}{c}{Parent Samples$^\dagger$} \\  
\hline 
\multicolumn{2}{c}{Mean}    & 12.6 & 8.14 & 1.29 & 22.45 & -1.26 & -1.96 & 7.86  \\
\multicolumn{2}{c}{Median}  & 9.2  & 7.79 & 0.94 & 22.81 & -1.81 & -1.99 & 7.80  \\
\multicolumn{2}{c}{Minimum} & 0.5  & 5.19 & 0.09 & 25.26 & -4.66 & -3.88 & 7.30  \\
\multicolumn{2}{c}{Maximum} & 71   & 8.85 & 5.60 & 18.66 & -0.27 & -0.01 & 8.74  

\enddata

\tablerefs{Data as reported in \cite{LTdata}. Original distance and metallicity references, respectively. DDO 69: \cite{Dol02}, \cite{vZ06}.  DDO 70: \cite{Sak04}, \cite{Kni05}.  DDO 75: \cite{Dol03}, \cite{Kni05}.  DDO 155: \cite{Tol95}, \cite{vZ06}.  WLM: \cite{Gier08}, \cite{Lee05}.} 
\tablecomments{General information about this galaxy sample, as reported by \cite{HunterHa,HunterUBV,LTdata}.  ${\mu_0}^V$ is the central $V-$band brightness.  R$_D$ is the disk scale length. SFR$^{FUV}$ is the star formation rate determined from L$_{FUV}$, and SFR$_D^{FUV}$ is that divided by $\pi$R$_D^2$.  Oxygen abundances for metallicities were determined from \hii\ regions. \\
$\dagger$ Parent sample statistics are drawn from the 40 \lt\ galaxies for the \hi\ masses and metallicities, and the 136-dwarf sample (which includes \lt) for the rest.} 

\end{deluxetable*}

This paper is organized as follows.  
In \S~\ref{sec:data} we describe our data treatment, in \S~\ref{sec:analysis} we discuss the various analysis techniques and formulations used in this study, in \S~\ref{sec:results} we detail the FIR flux ratios as well as qualitative features in the maps, and in \S~\ref{sec:discussion} we discuss our results in further detail before summarizing in \S~\ref{sec:conclusion}.


\begin{deluxetable*}{ clcccccc }

\tablecaption{ Observations \label{table:Obs}} 
\tablehead{ \colhead{Galaxy} & \colhead{Line} & \colhead{RA (J2000)} & \colhead{DEC (J2000)} & \colhead{OBSID} & \colhead{Date} & \colhead{Duration} & Field of View \\
			 & & (h m s) & (d m s) & & (YYYY-MM-DD) & (s) & (pc per side, 51\farcs5) }

\startdata
DDO 69  & \cii$_{\lambda 158}$  & 09 59 33.37 & +30 44 37.63 & 1342232282 & 2011-11-12 & 4612 & 200 \\
        & \oi$_{\lambda 63}$    & 09 59 33.32 & +30 44 36.18 & 1342232283 & 2011-11-12 & 8380 & \\
        & \oiii$_{\lambda 88}$  & 09 59 33.02 & +30 44 27.35 & 1342232284 & 2011-11-12 & 1787 & \\
        & \nii$_{\lambda 122}$  & 09 59 33.39 & +30 44 38.15 & 1342232312 & 2011-11-13 & 23,684 & \\ \\
        
DDO 70  & \cii$_{\lambda 158}$  & 09 59 58.20 & +05 19 40.07 & 1342233705 & 2011-12-06 & 4612 & 325 \\
        & \oi$_{\lambda 63}$    & 09 59 58.15 & +05 19 38.44 & 1342233706 & 2011-12-06 & 8380 &  \\
        & \oiii$_{\lambda 88}$  & 09 59 57.89 & +05 19 29.71 & 1342233707 & 2011-12-06 & 1787 &  \\
        & \nii$_{\lambda 122}$  & 09 59 58.19 & +05 19 39.83 & 1342233708 & 2011-12-06 & 23,684 &  \\ \\
        
DDO 75  & \cii$_{\lambda 158}$  & 10 11 06.13 & -04 42 23.29 & 1342232588 & 2011-11-21 & 4612 & 325  \\
        & \oi$_{\lambda 63}$    & 10 11 06.09 & -04 42 24.95 & 1342232589 & 2011-11-21 & 8380 &  \\
        & \oiii$_{\lambda 88}$  & \nodata  &  \nodata & \nodata & \nodata & \nodata \\
        & \nii$_{\lambda 122}$  & 10 11 06.13 & -04 42 23.55 & 1342232590 & 2011-11-21 & 23,684 &  \\ \\
        
DDO 155 & \cii$_{\lambda 158}$  & 12 58 40.10 & +14 12 56.86 & 1342236274 & 2012-01-03 & 4612 & 549  \\
        & \oi$_{\lambda 63}$    & 12 58 40.09 & +14 12 56.63 & 1342236275 & 2012-01-03 & 8380 &  \\
        & \oiii$_{\lambda 88}$  & 12 58 39.82 & +14 12 47.87 & 1342236276 & 2012-01-03 & 1787 &  \\
        & \nii$_{\lambda 122}$  & 12 58 40.13 & +14 12 58.02 & 1342236277 & 2012-01-04 & 23,684 &  \\ \\
        
WLM     & \cii$_{\lambda 158}$  & 00 02 01.62 & --15 27 51.42 & 1342236873 & 2012-01-08 & 4612  & 250 \\
        & \oi$_{\lambda 63}$    & 00 02 01.63 & --15 27 51.63 & 1342237491 & 2012-01-14 & 8380 &  \\
        & \oiii$_{\lambda 88}$  & 00 02 01.40 & --15 27 40.50 & 1342236874 & 2012-01-08 & 1787 &  \\
        & \nii$_{\lambda 122}$  & 00 02 01.63 & --15 27 51.87 & 1342236282 & 2012-01-04 & 23,684 &  \\
\enddata


\end{deluxetable*}

\begin{figure*}
\centering
\includegraphics{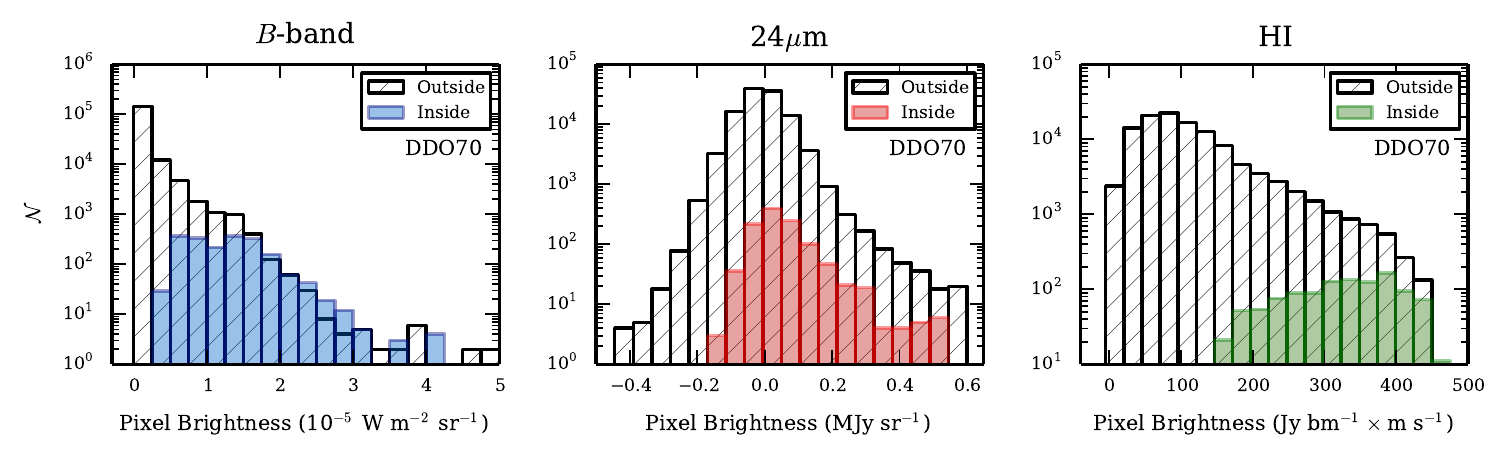}
\caption{Pixel brightness histograms for the $B-$band, 24$\mu$m, and \hi\ maps of DDO 70.  The values  inside the observed \herschel\ footprint are shown in color, while those outside of the footprint are shown in black and white.  The ``inside'' distributions of the $B-$band and 24$\mu$m maps are representative of the majority of the ``outside'' distributions.  \hi\ is quite extended in our galaxies, with low column densities at the outer edges (c.f. Figures~\ref{fig:zoom} \&~\ref{fig:zoomonline}).  Even though the \hi\ distributions are not equivalent, we sample the brighter non-edge emission in the galaxy fairly well.  The histograms for the other four galaxies in our sample show similar behavior to those presented here for DDO 70.}
\label{fig:insideoutside}
\end{figure*}

\pagebreak
\section{DATA}
\label{sec:data}

\subsection{Observations}
\label{sec:observations}

We used the PACS spectrometer aboard \herschel\ to observe the \cii\ 158$\mu$m, \oi\ 63$\mu$m, \oiii\ 88$\mu$m, and \nii\ 122 $\mu$m fine-structure lines in star-forming regions of five galaxies from the \lt\ sample: DDO 69, DDO 70, DDO 75, DDO 155, and WLM.  All four lines were observed for each galaxy, with the single exception of DDO 75, which had no \oiii\ observation.  A single star-forming complex was chosen for the center of the PACS footprint in each galaxy. The specific regions observed were selected as being most likely to have molecular gas, based on areas of enhanced far-ultraviolet (FUV) and H$\alpha$ -- indicating recent star formation -- as well as neutral hydrogen (\hi) emission.  Figures~\ref{fig:zoom} and ~\ref{fig:zoomonline} show the target regions in comparison to the \hi\ distribution.  A summary of the observations is given in Table ~\ref{table:Obs}.

Although the observed regions were selected based on likely sites of star formation, they are nevertheless representative of each host galaxy in several wavebands.  Figure~\ref{fig:insideoutside} shows $B-$band, 24$\mu$m, and \hi\ brightnesses of pixels inside and outside of the \herschel\ footprint for DDO 70.  The outer edges of each galaxy have many lines of sight with extremely low surface brightness, as expected, but the regions observed with \herschel\ are representative of the non-edge regions.  For a selection of global parameters of our galaxies compared to the other samples discussed in this work, see Table ~\ref{table:samplecomp}.

\begin{deluxetable}{ lccc }
\tablecaption{ Median Galaxian Properties of the Various Samples \label{table:samplecomp}} 
\tablehead{ \colhead{Sample} & \colhead{12+log(O/H)$^\dagger$} & \colhead{log$_{10}$ SFR} & \colhead{M$_*$} \\
                             &                         & (M$_\odot$ yr$^{-1}$) &  (10$^8$ M$_\odot$)   }
\startdata
Our Sample                 & 7.50  & -2.15 & 0.31 \\
\lt                        & 7.84  & -1.89 & 0.59 \\
DGS                        & 7.91  & -1.05 & 4.25 \\
\begin{tabular}{@{}c@{}}\citet{Brauher} \\ Irregulars\end{tabular} & 8.36  & 0.77 & 11.90 \\
\enddata
\tablecomments{ Metallicities for the \citet{Brauher} sample were determined from the L-Z relation described by \citet{Lamareille2004}.}
\end{deluxetable}

The targets were observed with a chopping/nodding raster pattern in the Open Time 1 (OT1) mission phase over a total of 53 hours.  Observed lines are not spectrally resolved, with the velocity FWHM ranging from roughly 90 km/s for the \oi\ 63$\mu$m line to 290 km/s for the \nii\ 122$\mu$m line. The other widths are 125 km/s and 240 km/s for \oiii\ 88$\mu$m and \cii\ 158$\mu$m, respectively.  The line maps in each galaxy were observed using 2$\times$2 rasters of 4\farcs5 steps. All observations were performed in the chop-nod mode with a large off-source chop throw of 6\arcmin, sufficiently far away from any star-forming regions associated with the target galaxy. The only source with possible contamination from chopping positions is WLM, due to its extended nature.  The \hi\ emission in WLM does extend out to just about 6\arcmin, however it is very tenuous that far out -- column densities drop by a factor of $\sim$20. The entire field of view covered for each galaxy is the sum of the PACS grid of 5$\times$5 spatial pixels (referred to as \textit{spaxels}, 9\farcs 4 on each side) plus the raster step, for a total of 51\farcs 5 per side.

\subsection{Reduction}
\label{sec:reduction}

\begin{figure}
\centering
    \begin{subfigure}[]{}
        \includegraphics[width=0.4\textwidth]{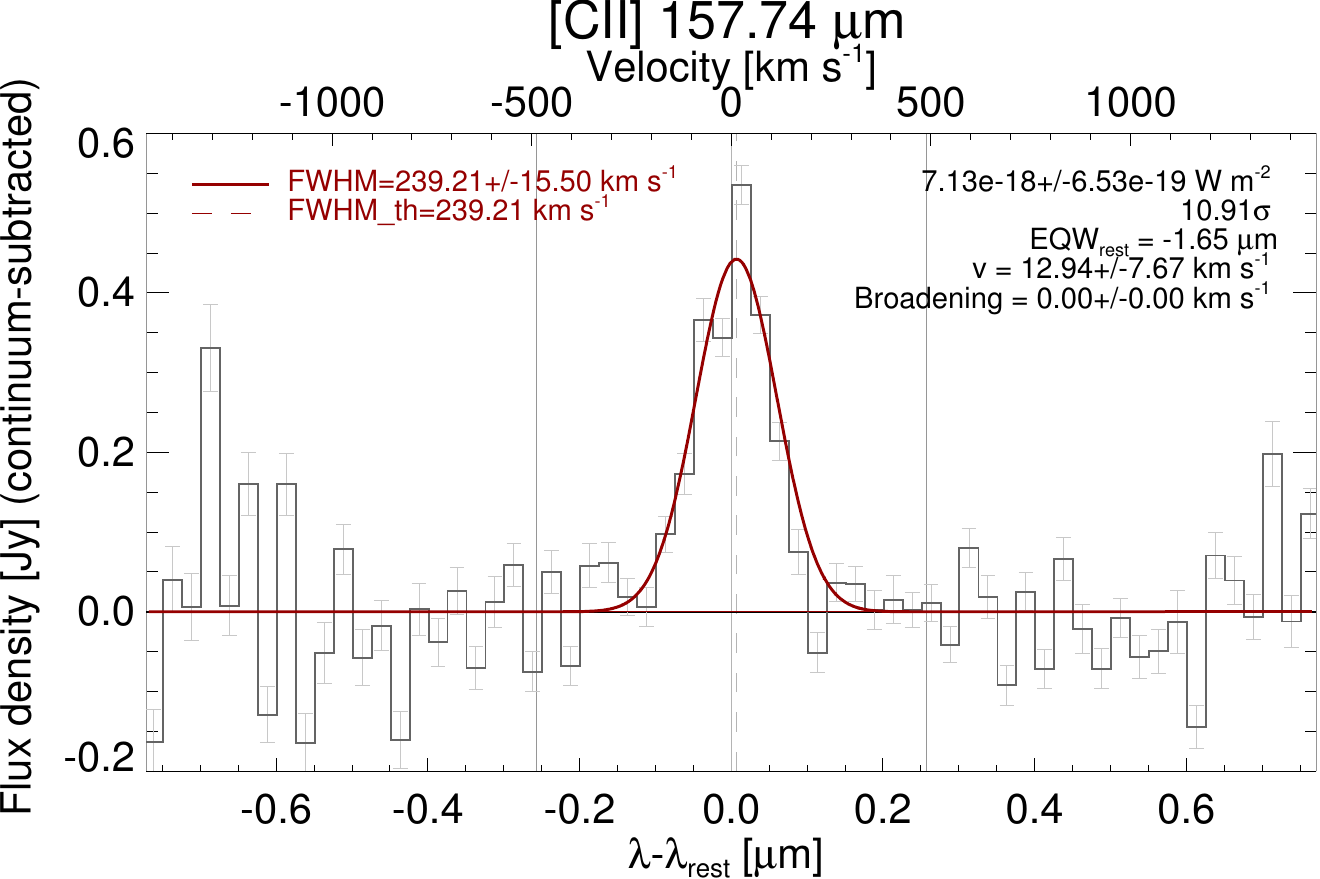}
    \end{subfigure}
    \\
    \begin{subfigure}[]{}
        \includegraphics[width=0.4\textwidth]{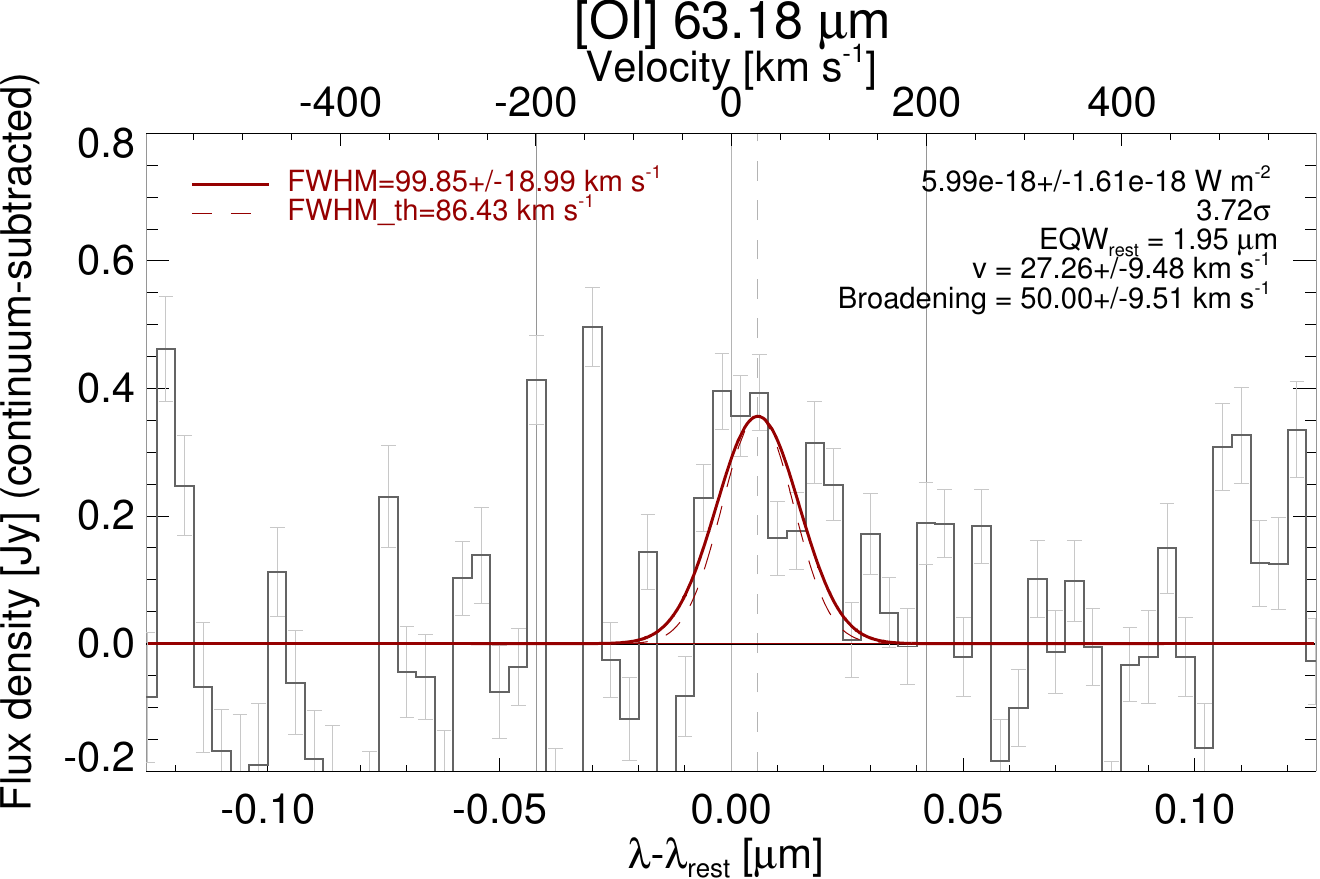}
    \end{subfigure}
    \\
    \begin{subfigure}[]{}
        \includegraphics[width=0.4\textwidth]{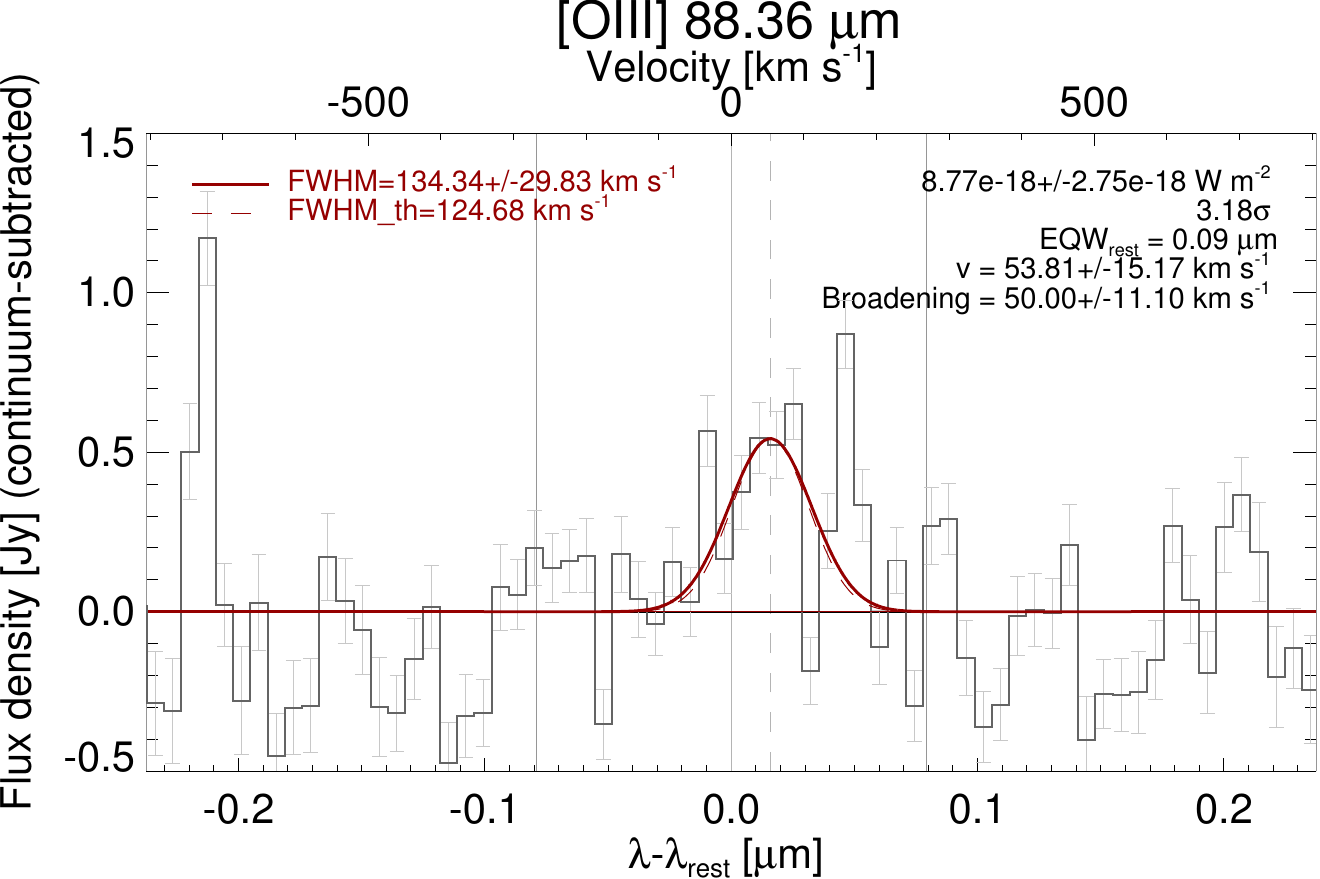}
    \end{subfigure}
    \\
    \begin{subfigure}[]{}
        \includegraphics[width=0.4\textwidth]{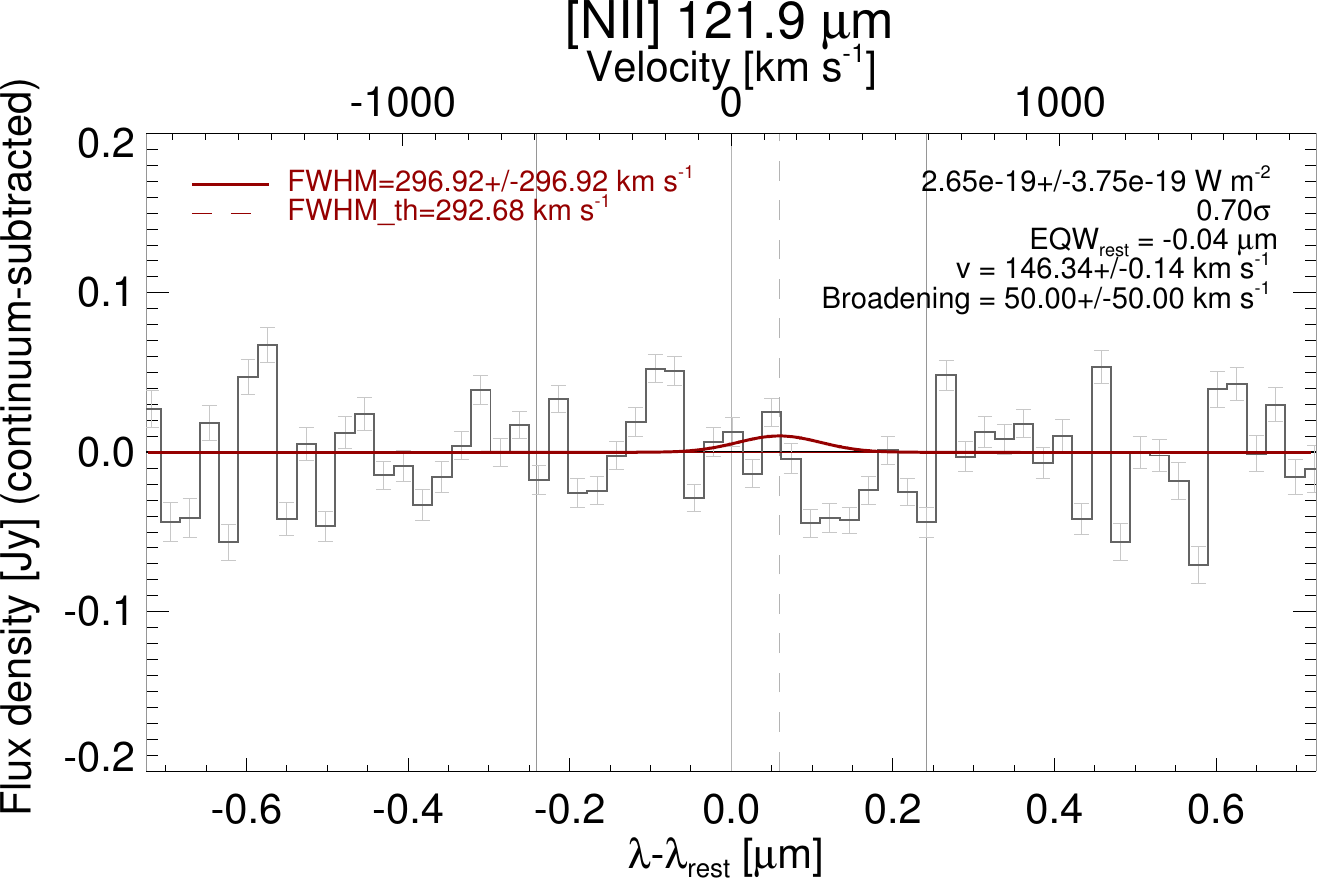}
    \end{subfigure}
\caption{Sample continuum-subtracted spectra from single rasters of the central spaxel in the PACSman fit procedure for DDO 155. We detect \cii, \oi, and some \oiii\ emission, but there is no cospatial \nii\ signal in this particular region.  Line fit values are noted within each plot. The line center velocity is denoted as v.  FWHM$_{th}$ is the spectral resolution line width, due to the thermal properties of the instrument.  Broadening is the excess line width beyond FWHM$_{th}$ in the fit, when added in quadrature. That is, FWHM$^2$ = FWHM$_{th}^2$+Broadening$^2$. This broadening is not likely physical.  EQW$_{rest}$ is the width in wavelength units for which a box defined from zero flux to the continuum level, and centered at the rest wavelength, has an area equal to that of the integrated flux.}
\label{fig:spectra}
\end{figure}

Basic data reduction was performed using the \herschel\ Interactive Processing Environment v11.1.0 \citep[HIPE;][]{Ott10} with calibration tree 56.  The HIPE ChopNod LineScan pipeline was used to calibrate flux and wavelength, as well as mask bad pixels -- those that are outliers, glitches, saturated, or noisy -- resulting in ``Level 1'' cubes.  We utilize two sets of final data products for our analysis: 2-dimensional integrated flux maps; and 3-dimensional data cubes, with the third axis containing the spectral information.  Although the lines are not spectrally resolved in the PACS data, the 3D data cubes are useful for combining spectra over spatially extended regions to draw out faint emission that would not be detected by 2D flux map integration alone.

The 2D images for each of the observed lines were created using the PACSman package \citep{Leb12}, which is a suite of utilities for line-fitting, mapping and analyzing \herschel\ data that have been calibrated and masked by HIPE. Normally within HIPE, a spectrum for each of the 25 spaxels is created by rebinning the calibrated and masked data ``cloud'' of up to several hundred thousand readouts in each spaxel.  From there, the user can perform line fits and integrate fluxes with their preferred tasks.  This process can average out much of the spatial and spectral information available in the data before fitting, such as minor dithers in the RA and DEC of nod positions and slight spatial dependences of spectral layers in the cubes, which can result in a reduction of the dynamic range of the final images.  In PACSman, by contrast,  Gaussian-plus-polynomial profiles are fit to the full data cloud of line-plus-continuum emission at each spaxel in each raster position, utilizing all available information, without prior rebinning.  Fits to some sample spectra of our observed fine-structure lines are shown in Figure~\ref{fig:spectra}.

We utilize the Monte Carlo option to robustly estimate the fit errors.  PACSman does this by iteratively adding random perturbations to the spaxel data cloud values and then fitting the perturbed data. First, the data for a spaxel are binned.  The error attributed to the individual points in a bin is the standard deviation of the values contained within.  Then this error is used as the dispersion ($\sigma$) in a Normal distribution to generate the perturbation for a given data point.  The final error is taken to be the standard deviation of the distribution of fit results from the perturbed flux arrays.  Tests on several maps determined that 400 iterations is the optimal number to produce stable and reasonable error estimates that do not differ significantly from those produced by higher numbers of iterations.

To produce the final maps, PACSman uses an approach to raster map projection similar to Drizzle \citep{FruchterHook}, where the rasters are projected onto a sub-spaxel grid of $\sim$3\arcsec pixels.  The value at each projected pixel is determined from the corresponding spaxel fraction, assuming uniform surface brightness in each spaxel.  Thus, each pixel in the final projection grid can correspond to several line fits, which are weighted and averaged to make the resulting 2D flux maps (see, for instance,  Figures~\ref{fig:zoom} and ~\ref{fig:zoomonline}).  

The 3D cubes are called \spc\ in the HIPE parlance, and are the final products from the pipeline. All of the rasters, initially organized in sets of 5$\times$5 spaxels, are projected onto a grid of smaller pixels.  Where PACSman projects fitted fluxes, however, HIPE projects the rebinned spectra directly, resulting in spatial maps of uniformly binned spectra.  We set the spatial \texttt{oversample} setting to 2 in HIPE, and the spectral \texttt{upsample} to 1 to produce our final data cubes -- the latter is particularly important for appropriate error estimation, as it ensures the spectral channels are independent from each other. 

PACSman can give more reliable fitted fluxes for given pixels, for the reasons mentioned above, and the superior dynamic range of the these maps aids in comparing different regions within the image.   However, the preservation of spectral information is crucial for some applications.  Pixel errors can be quite large for faint flux levels, so combining the spectra from the 3D cubes directly over an area of interest before fitting can provide a higher S/N ratio for the \textit{integrated} emission of an extended region.  While integrated fluxes of the bright \cii\ line are roughly the same using both data products, the 3D spectral cubes are better for characterizing the integrated emission from the other three lines in our galaxies.  To summarize: in our low surface-brightness regime, the PACSman maps provide better morphological and qualitative information for bright emission, while the 3D cubes give better integrated fluxes over multiple pixels.  The method of combining spectra is discussed in more detail in $\S$~\ref{sec:stackspec}.

\subsection{Ancillary Data}
\label{sec:ancillary}

We supplement the \herschel\ data with mid- and far-infrared continuum images from the \textit{Spitzer Space Telescope} at 24, 70, and 160 $\mu$m, as well as our PACS Photometry 100$\mu$m and 160$\mu$m maps, where available. We utilize \textit{Infrared Astronomical Satellite} (\iras) 60 and 100 $\mu$m measurements for comparison with previous studies.  We also have optical images in $V-$ and $B-$bands, H$\alpha$, and FUV, for comparison with features and luminosities seen in the infrared.  These maps also allow us to locate areas and quantify rates of star formation in our galaxies.

It is insightful to compare our line fluxes to integrated broadband infrared emission, as the ratios of line fluxes to broadband photometry measurements can be used with PDR models to determine physical parameters such as number densities and radiation field intensities.  We only use \spitzer\ and \herschel\ photometry to calculate TIR fluxes for our sample, though we use the \iras\ data with much coarser angular resolution to obtain globally averaged 60$\mu$m/100$\mu$m flux ratios for comparison with the literature. The various far infrared data sets used for these purposes are described below.
 
All five of our galaxies were mapped in 24, 70, and 160$\mu$m continuum with \spitzer\ and presented by \cite{Dale09} as part of the \spitzer\ Local Volume Legacy Survey.  We use these \spitzer\ maps to calculate global TIR values (discussed further in $\S$~\ref{sec:TIRvsFIR}).   Three of our five galaxies -- DDO 69, DDO 70, and DDO 75 -- were observed with the PACS and SPIRE photometers aboard \herschel\ during OT2.  The final maps were made with the Scanamorphos package \citep{scanamorphos}.  The full discussion of the \herschel\ photometry data reduction will be covered in our upcoming paper (Cigan et al., in prep.).   The resolution of these data is closely matched to the spectral observations, and so they are preferable to the coarser \spitzer\ measurements in calculations of TIR for comparisons with the spectral lines.  Since there is no \herschel\ photometry for DDO 155 or WLM, though, we primarily use the \spitzer\ maps for globally integrated TIR calculations to ensure a consistent source for all five galaxies, and use the \herschel\ photometry maps where available as supplementary data.

Unresolved \iras\ 60$\mu$m and 100$\mu$m fluxes were available for all of our galaxies: DDO 69 \citep{Tacconi87}, DDO 70 \citep{Lisenfeld07}, DDO 75 \citep{Moshir90}, DDO 155 \citep{Melisse94}, and WLM \citep{Rice88}.  We only use the ratio of unresolved IRAS 60$\mu$m and 100$\mu$m fluxes to compare our global line ratios with those of our reference sample discussed below; we do not use the \iras\ data to compute broadband TIR fluxes for our galaxies.

We use the $B-$band images obtained by \citetalias{HunterUBV} to determine TIR/B ratios, which are a measure of the starlight reprocessed by dust. The total uncertainties reported combine the statistical errors of integration and the overall systematic errors. GALEX FUV maps come from \cite{HunterFUV}, with the exception of DDO 155, which comes from \cite{Dale09}.

We use the \lt\ maps of the neutral hydrogen (\hi) content in these galaxies, which are based on NRAO\footnote{The National Radio Astronomy Observatory is a facility of the National Science Foundation operated under cooperative agreement by Associated Universities, Inc.} Very Large Array (VLA) observations in B-, C- and D-configurations.  The \hi\ maps have a common angular resolution of 6\arcsec.  The data were presented by \cite{LTdata} and can be downloaded from the \lt\ NRAO science web page \footnote{\url{http://science.nrao.edu/science/surveys/littlethings}}.

In our line ratio plots, we make use of the data points presented by \citet[hereafter \citetalias{HunterISO}]{HunterISO} for observations of other \lt\ galaxies with \iso\ and \iras, as well as data for a variety of other galaxy types presented by \citet[hereafter \citetalias{Brauher}]{Brauher} and references therein.  The IRAS 60/100$\mu$m values for DDO 69 and DDO 155 were not reported in those studies, and we take their values from the following additional sources.  Fluxes and uncertainties for DDO 69 60$\mu$m, DDO 69 100$\mu$m, and DDO 155 60$\mu$m measurements come from \cite{Tacconi87}.  The same authors only report an upper limit for the 100$\mu$m flux in DDO 155, while \cite{Melisse94} report a flux detection without an accompanying uncertainty. For this source, we use the published detection as the flux, and take the 1$\sigma$ upper limit as the uncertainty.

\subsection{Convolving the Data}
\label{sec:smoothingdata}

When comparing data with different resolutions, it is essential to convolve the pertinent maps to a common resolution. Comparisons of the PACS \oi\ with the \oiii\ maps require no further processing, as they have a common beam size of $\sim$9\farcs5 FWHM \citep{Pog10}, though comparison with \cii\ maps requires smoothing to the 157$\mu$m beam size of 11\farcs5.  ``Beam size'' refers to the resolution element, or the PSF Full-Width-at-Half-Maximum (FWHM).  

The \spitzer\ maps have different resolutions in the different bands.  We use the values reported by \cite{Aniano11} for the \spitzer\  beams: 6\farcs 5 for 24$\mu$m; 18\farcs 7 for 70$\mu$m; 38\farcs 8 for 160$\mu$m.  Table~\ref{table:instrumentparams} summarizes the beam sizes of each instrument.  The size of the MIPS 160$\mu$m beam is similar to the entire PACS field of view.  Any small resolved aperture within a \herschel\ map will thus be on the order of (or smaller than) the MIPS$_{160}$ beam, so we make no spatially resolved comparison between the two instruments on those scales.  All \spitzer\ maps are convolved to the common beam size of 38\farcs8 FWHM before they are combined to derive values for the TIR continuum.

\begin{deluxetable}{ llrr }
\tablecaption{ \herschel\ and \spitzer\ Instrument Parameters \label{table:instrumentparams}} 
\tablehead{ \colhead{Instrument} & \colhead{Band} & \colhead{Beam Size} & \colhead{Calibration} \\
                                 &                &  (FWHM)             &  Uncertainty             }

\startdata
PACS & \cii  & 11\farcs5 & 12\% \\
PACS & \oi   & 9\farcs5 & 11\% \\
PACS & \oiii & 9\farcs5 & 12\% \\
PACS & \nii  & 10\farcs0 & 12\% \\
\\
MIPS & 24$\mu$m  & 6\farcs5   & 4\% \\
MIPS & 70$\mu$m  & 18\farcs7  & 10\% \\
MIPS & 160$\mu$m & 38\farcs8  & 12\% \\
\enddata
\tablerefs{ Beam sizes come from \cite{Pog10} (PACS) and \cite{Aniano11} (\spitzer\ MIPS). Instrumental calibration uncertainties are as follows.  PACS: The PACS Observers' Manual \footnote{\url{http://herschel.esac.esa.int/Docs/PACS/html/pacs_om.html}}; MIPS 24$\mu$m: \cite{Eng07}; MIPS 70$\mu$m: \cite{Gor07}; MIPS 160$\mu$m: \cite{Sta07}.  }
\end{deluxetable}

\section{ANALYSIS}
\label{sec:analysis}

\subsection{Outline}
\label{sec:analysisplan}

Our goal is to study the physical properties of the galaxies we observed.  We first make note of the general qualitative properties apparent in the maps produced by the standard data reduction as described in $\S$~\ref{sec:data}.  Then we quantify the line flux in each map by combining spectra from several pixels, fit a profile to the combined spectrum, and examine the appropriate uncertainties that apply to our measurements.  Next, we determine the broadband infrared flux over the spatial scale of the PACS line footprints, in order to compare the various line emission fluxes to the integrated continuum.   Line and broadband fluxes are additionally determined for resolved regions within maps, where possible, to study our systems spatially.  Finally, we discuss the results of our flux integrations within this sample of five galaxies, and compare them with previously published results.  To quantify the fundamental gas properties in our targets, we pursue more detailed modeling in upcoming papers to obtain dust masses from the broadband photometry, and combine the photometry and spectroscopy in PDR models to obtain densities and radiation field estimates.

\subsection{Qualitative Remarks About Map Fluxes}
\label{sec:qualitativeremarks}

There are several qualitative characteristics of our line maps that immediately stand out.  All observations exhibit, in general, quite low surface brightnesses.  Typical ``detection'' S/N over a region is $\sim$3-5 for \oi, \oiii, and \nii, with a typical brightness of $\sim10^{-9}$ W m$^{-2}$ sr$^{-1}$.  \cii\ emission is detected at high significance in all five galaxies, though -- maximum \textit{pixel} S/N values are $\sim$10-15, and even pixels near the map edges often have S/N$>$5. Thus we are certainly not recovering all of the \cii\ flux in our galaxies.  While some targets show enhanced \cii\ flux in certain resolved regions, there is generally emission extended over the majority of each image.  The \oiii\ emission is, on average, concentrated in a few pockets. The \nii\ detections in our maps are marginal ($\sim 3\sigma$), though this result is not entirely unexpected, since \nii\ is usually a factor of 10 fainter than the other FIR lines in galaxies.   

There are often bright spots at the edges of the images.  An inspection of these spots in the data cubes reveals that this is usually not genuine emission; reduced coverage can skew the averages, but the errors are correspondingly high.  We generally ignore the noisy outer edges of the maps and consider only the bright regions in the interior.  One notable exception is the southwest corner of the DDO 155 \cii\ map (see Figure~\ref{fig:zoomonline}), which shows some emission that is cospatial with a region of high-density neutral hydrogen, $V-$band, FUV, H$\alpha$, and 160$\mu$m signal.  DDO 69 shows a slight knot of \nii\ in the southwest corner of the map that is cospatial with 160$\mu$m emission and between clumps of V, FUV, and H$\alpha$.  However, only one raster registered a detection of 3.7$\sigma$ amid much noise, while it is not detected in the other rasters.

\subsection{Determining Integration Regions}
\label{sec:stackspec}

\begin{figure*}
\includegraphics[width=\textwidth]{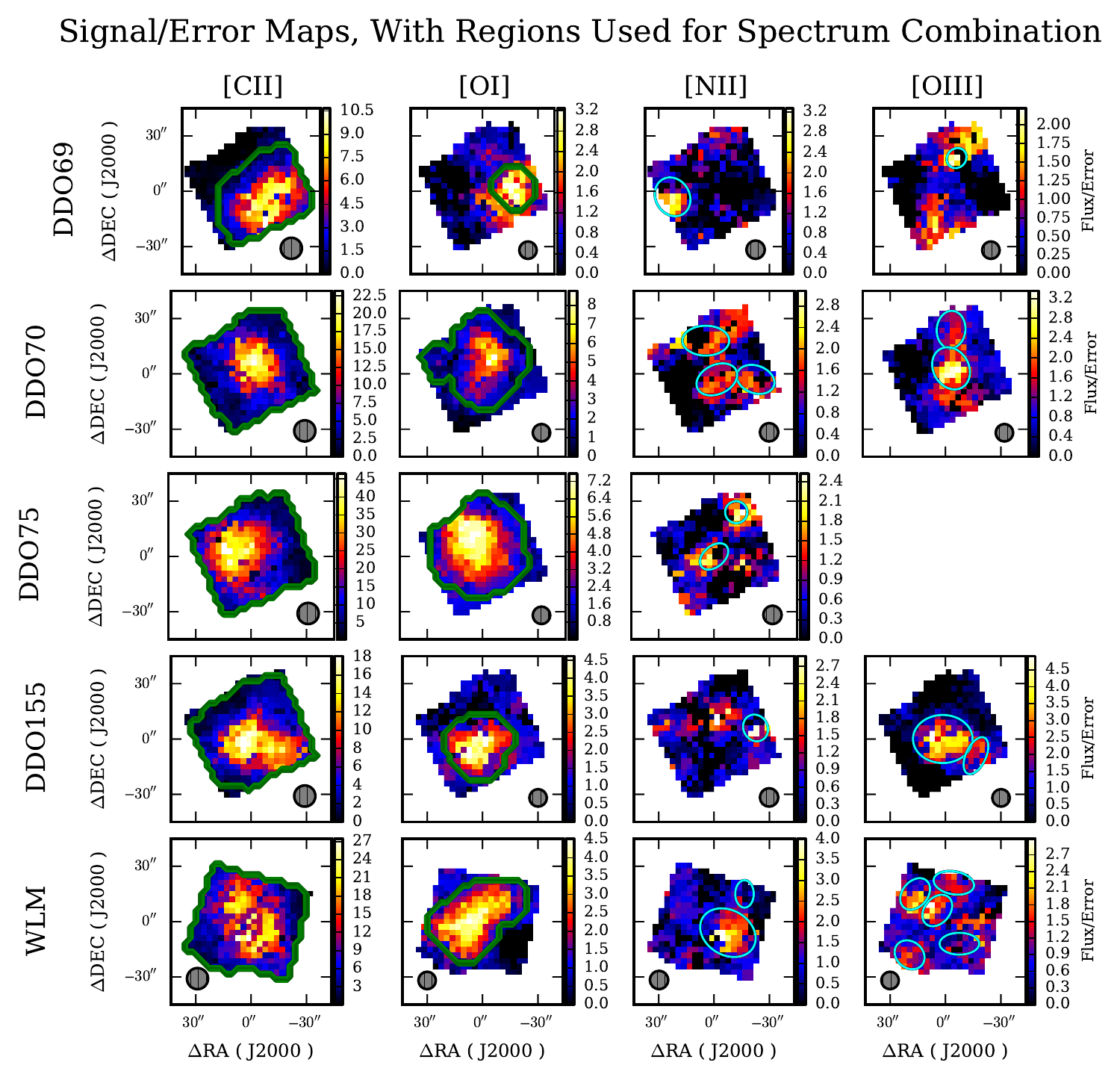}
\caption{Maps of signal to noise (fitted flux over fitted error) for our galaxies, with regions indicated for combining spectra.  Colorscale values indicate the significance ($\sigma$) level of the fits for spectra summed from a region centered on that pixel, with a diameter of two beam widths.  The green contours, used for the broadly detected \cii\ and \oi\ emission, outline the contiguous region in each map where S/N$>$3 (based on the 2-beam-width integration kernel).  The cyan ellipses in the \oiii\ and \nii\ maps are the regions used for spectrum aggregation, based on signal found by manual inspection of the 3D spectral cubes: apertures of varying sizes and shapes that contain S/N$\sim$3 profiles, based on the brightest regions in the moving-integration maps.  Note that the cyan ellipses do not exactly follow S/N peaks - that is, some regions of manually-identified signal do not overlap exactly with the bright spots from the integrated flux maps.  This is especially true for \nii, where the emission is particularly faint.}
\label{fig:SEstackregs}
\end{figure*}

\begin{figure*}
\centering
\includegraphics[width=\textwidth]{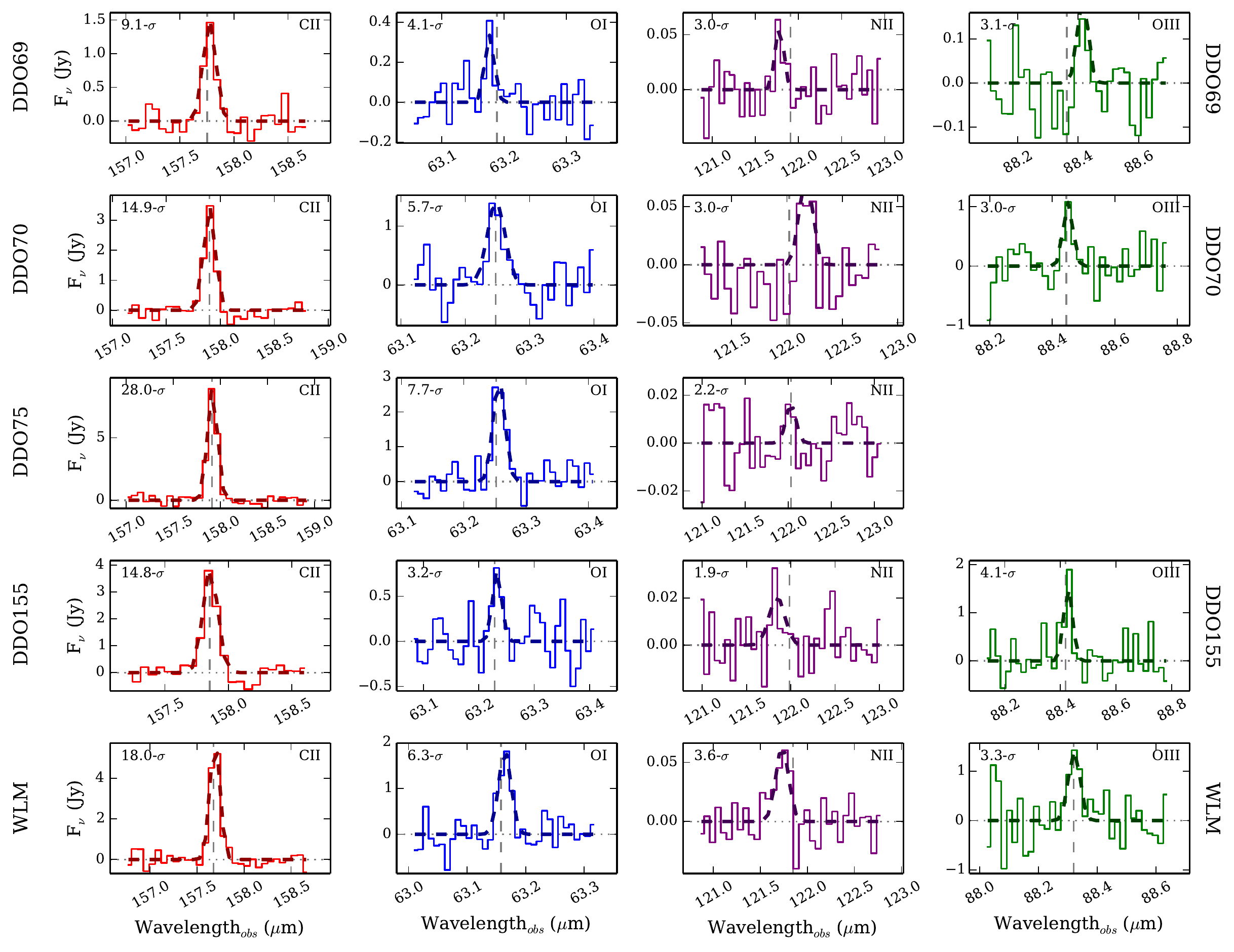}
\caption{Spectra from the regions described in Figure~\ref{fig:SEstackregs} are combined to give a global measure of all observed flux. The vertical dashed grey line indicates the expected line center location, based on the local \hi\ velocity of each galaxy. }
\label{fig:stackspec}
\end{figure*}

All of the PACS line fluxes we present in this work are determined by combining spectra over some spatial region.  While the \cii\ line is well-detected in all our galaxies, the other line maps give many non-detections when the entire cube is combined, especially for \nii\ and \oiii.  Inspecting these spectral maps by eye shows that there are indeed regions of detectable flux in many of these targets.

To address this, we created masks to restrict integration regions to areas with true emission. First, we create signal-to-noise maps based on specified integration kernels: circular boxcar functions using diameters of one beam FWHM, two beam widths, etc.  At every pixel position, the spectra enclosed in the adopted aperture size are combined and and fitted with a composite Gaussian and polynominal profile to determine the flux and error centered at that location.  The resulting maps, shown in Figure~\ref{fig:SEstackregs}, therefore give the detection significance of flux enclosed in the given aperture size centered at any given position.  Maps using different kernel sizes will show different views of the emission in a target, depending on how diffuse the emission regions are. A moving aperture of two beam widths is a good compromise that shows the majority of emission on the different scales without losing too much resolution, so the maps presented here use this kernel.  For bright and widespread emission, these maps are excellent masks for spectrum aggregation.  For our global measurements of \cii\ and \oi\ flux over each footprint, we combine all spectra within regions where the ``moving aperture'' signal to noise is greater than 3.

\begin{figure*}
\includegraphics[width=\textwidth]{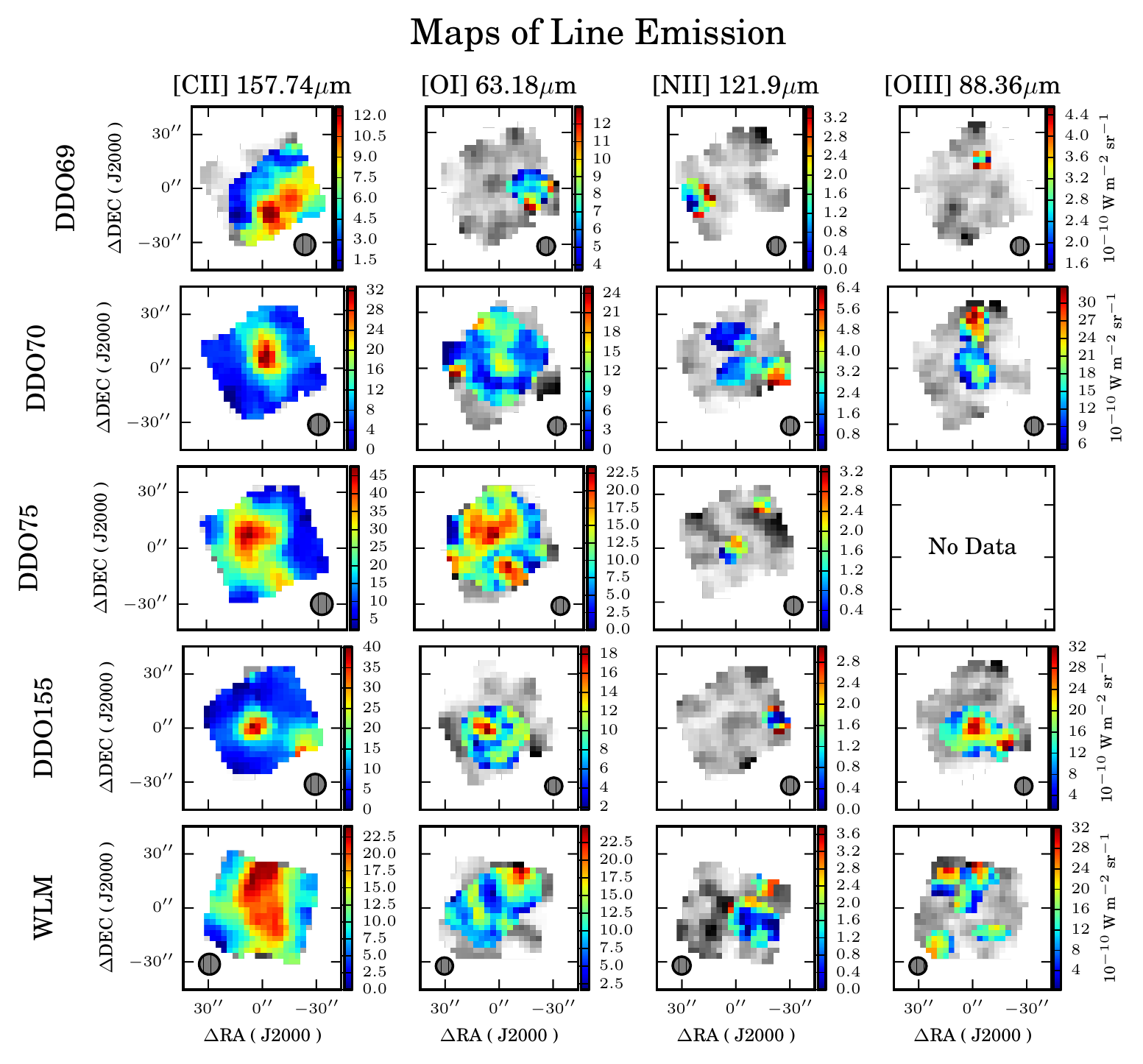}
\caption{PACSman brightness maps of the four observed lines in each of our five galaxies.  Fluxes in the regions that registered detections (S/N$>$3) are shown in color, while fluxes from areas of non-detections are shown in grayscale.  The flux (color) scale is different for each map.  The mask regions are the same as those detailed in Figure~\ref{fig:SEstackregs} and $\S$~\ref{sec:stackspec}: for \cii\ and \oi, S/N values are determined from the moving integration kernel maps, while S/N regions for \oiii\ and \nii\ were determined by manual inspection of the spectral data cubes.  The hatched circle at the bottom of each map shows the PACS beam size of the given line.}
\label{fig:coverage}
\end{figure*}

Regarding our \oiii\ and \nii\ maps, we manually identify individual regions where there are traces of genuine emission, based on our S/N maps.  Combining spectra from these hand-selected regions (shown by cyan ellipses in Figure~\ref{fig:SEstackregs})  results in final spectra with less noise. The resulting spectra aggregated from the regions defined in Figure~\ref{fig:SEstackregs} for each map are shown in Figure~\ref{fig:stackspec}.  The PACS flux maps showing the final integration regions are presented in Figure~\ref{fig:coverage}.  It should be noted that many of the regions with detectable faint signal are offset from the brightest regions in the PACSman maps.  Each region is no smaller than the limiting beam size for a given map.  Aperture corrections for the smallest regions, such as \oiii\ in DDO 69, are on the order of 20\%. They rapidly drop down to $\sim$2\% for apertures 20\arcsec\ wide, though, and are less than 1\% for footprint-scale regions.

When computing ratios of line fluxes, the apertures are matched according the intersection of the emission indicated in Figure~\ref{fig:SEstackregs}: the more prominent line is integrated over the area of the less prominent line. The order of selection for matching is \nii , \oiii , then \oi . For example, the fluxes used to compute the \nii/\cii\ both use the \nii\ integration regions. The optimized \nii\ and \oiii\ regions recover detections in most cases where the sums over the entire FOV do not -- the fluxes differ by $\lesssim$30\%, but the errors are much smaller.  They are appropriate for comparisons with low-resolution data in the literature, such as the reference sample of \iso\ measurements from \cite{Brauher}, but not for comparisons on scales smaller than the \herschel\ footprint because the \nii\ and \oiii\ apertures in particular are often offset with different sizes.

\subsection{Profile Fits and Uncertainties}
\label{sec:uncertainties}

Raw spectra from a given integration region are combined without prior continuum subtraction.  Baselines of individual spectra do not typically exhibit systematic patterns over beam-sized scales, which would otherwise be amplified by this method.  Line fluxes for a given spectrum are determined from the area under a Gaussian + polynomial profile: 

$f(\lambda) = A \ e^{\frac{1}{2} \left(\frac{\lambda-\lambda_c}{\sigma} \right)^2} + a \lambda^2 + b \lambda + c $.  

The resulting flux after subtracting the polynomial continuum response is thus given by $\mathrm{Flux}_{fit} = \sqrt{2 \pi} A \sigma \ $.  We use the python package \texttt{lmfit} for fitting, which is an implementation of the Levenberg-Marquart nonlinear least-squares algorithm.  \texttt{lmfit} calculates numerical derivatives with \texttt{MINPACK}'s \texttt{lmdif} function, as implemented by \texttt{scipy.optimize.leastsq}.  

Since our targets are extended with extremely low surface brightnesses, we rigorously examined the uncertainties of our measurements to ensure that they are reliably characterized. The two types of error we discuss below are the statistical error in the spectral fits and the general instrumental uncertainty of order 12\%.

For combined spectra, the fit error is the formal uncertainty of the area under the Gaussian  with independently fitted parameters $A$, $\lambda_c$, and $\sigma$: 

$\Delta_{\mathrm{Flux}_{fit}} = \mathrm{Flux}_{fit} \cdot \sqrt{ \frac{\epsilon_{A}^2}{A^2}  +  \frac{\epsilon_{\sigma}^2}{\sigma^2}  + 2 \frac{\epsilon_{A,\sigma}^2}{A \cdot \sigma}} \ ,\\ $ 
where $\epsilon$ has been used for the variances $\epsilon_{A}^2$, $\epsilon_{\sigma}^2$, and covariance $\epsilon_{A,\sigma}^2$ so as not to confuse it with the notation for the gaussian dispersion $\sigma$. 

The variances and covariances of the fitted parameters are determined automatically within the fitting routine: the standard errors come from the covariance matrix (C.M.) of the fit, which is calculated from the Jacobian of the system of equations: C.M. = $(J^TJ)^{-1}$.  The diagonal elements of the covariance matrix are the parameter variances (typically denoted as $\sigma^2$, but given as $\epsilon^2$ above), and the off-diagonal elements yield the covariances between parameters.  This assumes that all channels in the spectrum are independent of each other (\texttt{upsample}=1 in HIPE), otherwise the uncertainty will be underestimated.  Typical values for the errors on the fitted fluxes are around 10\%.  

The pixel error values used for the S/N ratios in the PACSman maps, by comparison, come from the fit uncertainties in each spaxel, which have been projected onto the pixel grid (see \cite{Leb12} for additional details on error handling within PACSman).

When comparing measurements from different instruments, we must include the uncertainty arising from overall instrumental effects.  The statistical uncertainties described above and those due to the system as a whole are added in quadrature.  The systematic uncertainties for the PACS Spectrometer and the \spitzer\ MIPS bands are summarized in Table ~\ref{table:instrumentparams}.  The sum total uncertainty for a specific integration region combines the summation (fit) error and the systematic error, which we compute as $\Delta_{\mathrm{tot}} = \sqrt{ \Delta_{fit}^2 +\Delta_{sys}^2}$.

\subsection{TIR Fluxes}
\label{sec:TIRvsFIR}

We wish to compare our line emission with the infrared continuum, to understand how much of the total infrared energy each line comprises, and to compare our sample with other galaxies.  However, some care must be taken with definitions of ``Far''- and ``Total''-Infrared emission (FIR, TIR, respectively), as there is some overlap in the literature.  In general, FIR is taken to be the range from several to some hundreds of microns, while TIR covers the range from several to a roughly a thousand microns. 

FIR has traditionally been preferred over TIR for characterizing the infrared continuum, for various reasons including the lack of instruments able to sample the longer wavelengths of the IR spectrum.  This was reasonable because observation targets were typically bright with thermal radiation that covered the FIR ranges well.  However, the narrow band of the canonical FIR (40-120$\mu$m) can be a poor match for the SEDs of cooler objects -- a simple blackbody spectrum at 25K peaks at 147$\mu$m, outside of the range. TIR also includes some emission from particles such as PAHs at wavelengths shorter than the FIR cutoff.  The SEDs of our systems peak around 100$\mu$m, but some of the objects in the Dwarf Galaxy Survey \citep[DGS;][]{DGS} peak below 40$\mu$m \citep[see][]{RemyRuyer13}, making FIR a particularly poor choice for IR characterization.  For these reasons, we use TIR in our analysis to better represent the majority of the emission in the infrared continuum.  

We determine TIR luminosities for our five galaxies from a combination of \spitzer\ MIPS measurements at 24, 70, and 160 $\mu$m. We also utilize TIR calculated from PACS 100$\mu$m for our three galaxies with \herschel\ photometry data -- DDO 69, DDO 70, and DDO75.  Since we only have \herschel\ photometry data for 3 of our five galaxies, we primarily use the \spitzer -based TIR for footprint-scale measurements to ensure a consistent source for the sample, and use the \herschel -based TIR in $\S$~\ref{sec:resolvedscales} where we analyze the relation of line emission and TIR on a spatially resolved basis.  When comparing \herschel\ line fluxes to TIR$_{Spitzer}$, the TIR values are for the entire PACS field of view due to the poor MIPS 160$\mu$m resolution, while the superior resolution of TIR$_{Herschel}$ allows for the appropriately matched apertures.  We exploit the higher spatial resolution of the PACS photometry for a resolved comparison of line ratios in $\S$~\ref{sec:regioncomp} for those three galaxies.  The comparison sample from \citetalias{Brauher} used FIR fluxes calculated from older IRAS measurements. We calculate the TIR fluxes for their sample directly, instead of applying a TIR/FIR ratio.  Table ~\ref{table:TIRsources} summarizes the data sources used to calculate TIR for the various samples discussed in this work.

\begin{deluxetable}{ lcc }
\tablecaption{ TIR Data Sources \label{table:TIRsources}} 
\tablehead{ \colhead{Sample} & \colhead{TIR Data Source}   }

\startdata
DDO 69, DDO 70, DDO 75, DDO 155, WLM    &  \spitzer\    \\ 
\\
DDO 69, DDO 70, DDO 75 $\ \ $(resolved) &  \herschel\   \\
\\
Reference Galaxies \citep{Brauher} &  \iras\
\enddata
\tablecomments{\herschel\ photometry is only available for three of our five galaxies.  \iras\ measurements are only used in TIR calculations for the reference sample, not for our galaxies in this work.  }
\end{deluxetable}

\citetalias{HunterISO} and \cite{Dale09}, following \cite{Dale02}, estimate the TIR from 3-1100$\mu$m as  $f(\mathrm{TIR})_{IRAS} = 2.403(\nu f_\nu)_{25 \mu m} - 0.2454(\nu f_\nu)_{60 \mu m} + 1.6381(\nu f_\nu)_{100 \mu m}$ and $f(\mathrm{TIR})_{MIPS} = 1.559(\nu f_\nu)_{24 \mu m} + 0.7686(\nu f_\nu)_{70 \mu m} + 1.347(\nu f_\nu)_{160 \mu m}$.  See \S 5.2 of \cite{Dale09} for further discussion.
\cite{DraineLi} add IRAC 8$\mu$m to the MIPS 24, 70, and 160 micron data to estimate the TIR luminosity.  We only have 8$\mu$m maps for DDO 69 and DDO 155, though, so in the interest of consistency, we use the definition of $f(\mathrm{TIR})_{MIPS}$ from \cite{Dale02} for all five galaxies.  
 
For our sources with \herschel\ photometry (DDO 69, DDO 70, DDO 75), we also estimate TIR using the prescription of \cite{Galametz13} based on PACS 100$\mu$m: log(L$_\mathrm{TIR}) = (1.0 \times) \ \mathrm{log}( \nu \mathrm{L}_{\nu \, 100})+0.256$,  where L$_\mathrm{TIR}$ and $\nu \mathrm{L}_\nu$ are in units of L$_\sun$. They demonstrated that this particular single-band formulation of TIR was a reliable estimator for most galaxies.

The uncertainties in the fluxes of each \spitzer\ and PACS photometry band were determined by comparison with off-source regions.  Each integration aperture was placed at several (6-8) locations in the map that were free of signal, yielding off-source fluxes for identical aperture sizes and shapes. The uncertainty of the original flux $\Delta f_\nu$ in a given map is then taken to be the standard deviation of these off-source measurements. The overall uncertainty in TIR is then derived according to standard error propagation.

\section{RESULTS}
\label{sec:results}

\subsection{Flux Integrations}
\label{sec:fluxintresults}

The \cii\ line is detected with high significance ($\sim$9-28$\sigma$ integrated) in all our targets, and is reliably detected along almost all lines of sight.  Even pixels at the edges of the maps generally have S/N$\sim$5, which means we are almost certainly missing some of the extended \cii\ in each galaxy.  Indeed, none of the footprints span the entirety of their target galaxies.  We note that there is generally not widespread (spanning multiple beam widths) reliably-detected \oi, \oiii, and \nii\ line emission in our galaxies, though \oi\ has $\sim 10\sigma$ integrated detections in three of our five galaxies over some moderately-sized regions.  

The \nii\ detection in WLM has the highest statistical significance in our sample, at 3.6$\sigma$.  The emission has an offset of roughly -260 km\,s$^{-1}$ from the galaxy's bulk velocity, however, which complicates matters.  This falls outside of the $1\,\sigma$ uncertainty set by the instrumental resolution and fit errors -- the \nii\ half-width at half-maximum is 145 km\,s$^{-1}$, and typical uncertainties on fitted velocities are no more than 10--20\% of that. It is conceivable that some of the velocity offset could be due to an effect discussed in the PACS Observers' Manual, where a source that is spatially offset from a spaxel's center can lead to offsets on the diffraction grating.  This only applies to emission that is compact with respect to the spaxels, and the WLM \nii\ emission is right at the edge of that scale.  This effect can only reasonably account for offsets of around 0.03$\mu$m for \nii, or one third of the roughly 0.1$\mu$m shift we see. One physical explanation could be that it may be due to emission from the ejecta of a supernova remnant, though nothing appears at those coordinates in the \lt\ radio continuum maps (Kitchener et al., submitted).  Regardless, the \cii, \oi, and \oiii\ emission in WLM is centered at the galaxy's recessional velocity, which means that these lines might not originate in the same physical location as the \nii.

Because we optimized the total integrated flux in each map in our image-total fluxes using the methods outlined in $\S$~\ref{sec:stackspec}, the integration regions in the \oiii\ and \nii\ maps are often smaller than those for \cii .  Using the same integration regions as \cii\ for these maps, however, does not increase the recovered fluxes in any map because of the addition of noisy spectra.  Furthermore, the S/N maps (e.g., Figure~\ref{fig:SEstackregs}) produced with the moving integration kernels determined that any flux in the areas external to our final defined regions were not bright enough to be detected at 3$\sigma$ significance over scales from 1-3 beam widths.  This means there is almost certainly not much missed gas that is near the surface brightness of our detections, and the fluxes we report are well-constrained.

\subsection{Flux Ratios}
\label{sec:ratios}

\begin{deluxetable*}{ lccccc }
\tablecaption{ Integrated Line Fluxes and Ratios \label{table:lineratios}} 
\tablehead{ \colhead{Quantity} & \colhead{DDO 69} & \colhead{DDO 70} & \colhead{DDO 75} & \colhead{DDO 155} & \colhead{WLM} }
 
\startdata
\\ \multicolumn{6}{c}{Combination of Spectra from All Pixels} \\ \hline \\ 
f$_{\text{\cii} \lambda 158}$ (10$^{-17}$ W m$^{-2}$) 	& $2.5 \pm 0.4$ & $5.4 \pm 0.5$ & $14.3 \pm 0.5$ & $7.6 \pm 1.2$ & $9.1 \pm 0.5$  \\ 
f$_{\text{\oi} \lambda 63}$ (10$^{-17}$ W m$^{-2}$)   	& ( 2.3 ) & $2.8 \pm 0.7$ & $5.3 \pm 0.8$ & ( 3.2 ) & $3.8 \pm 1.1$  \\ 
f$_{\text{\oiii} \lambda 88}$ (10$^{-17}$ W m$^{-2}$)	& ( 7.0 ) & ( 4.5 ) & \nodata         & ( 3.9 ) & ( 4.2 )  \\ 
f$_{\text{\nii} \lambda 122}$ (10$^{-17}$ W m$^{-2}$)	& ( 0.8 ) & ( 1.0 ) & ( 0.8 ) & ( 1.1 ) & ( 0.7 )  \\ \\ 
 
\multicolumn{6}{c}{Spectra Combined from Selected Regions With Signal} \\ \hline \\ 
f$_{\text{\cii} \lambda 158}$ (10$^{-17}$ W m$^{-2}$) 	& $2.7 \pm 0.3$ & $5.6 \pm 0.3$ & $14.1 \pm 0.4$ & $6.8 \pm 0.6$ & $9.2 \pm 0.5$  \\ 
f$_{\text{\oi} \lambda 63}$ (10$^{-17}$ W m$^{-2}$)   	& $0.4 \pm 0.1$ & $2.5 \pm 0.7$ & $5.3 \pm 0.7$ & $1.2 \pm 0.4$ & $3.4 \pm 0.6$  \\ 
f$_{\text{\oiii} \lambda 88}$ (10$^{-17}$ W m$^{-2}$)	& $0.3 \pm 0.1$ & $1.4 \pm 0.4$ & \nodata         & $2.1 \pm 0.4$ & $2.8 \pm 0.6$  \\ 
f$_{\text{\nii} \lambda 122}$ (10$^{-17}$ W m$^{-2}$)	& $0.12 \pm 0.03$ & $0.24 \pm 0.07$ & ( 0.10 ) & ( 0.06 ) & $0.20 \pm 0.06$  \\ \\ 
 
f$_{\text{\oi}_{\text{convolved-[CII]}}}$ (10$^{-17}$ W m$^{-2}$)  * & $0.4 \pm 0.1$ & $3.4 \pm 0.7$ & $5.1 \pm 0.6$ & $1.2 \pm 0.4$ & $3.6 \pm 0.6$ \\ 
f$_{\text{\oiii}_{\text{convolved-[CII]}}}$ (10$^{-17}$ W m$^{-2}$)  * & ( 0.5 ) & $1.3 \pm 0.4$ & \nodata         & $2.1 \pm 0.4$ & $2.6 \pm 0.6$ \\ 
f$_{\text{\nii}_{\text{convolved-[CII]}}}$ (10$^{-17}$ W m$^{-2}$) * & $0.12 \pm 0.04$ & $0.23 \pm 0.07$ & ( 0.07 ) & ( 0.08 ) & $0.21 \pm 0.06$ \\ 
 \\ 
\textbf{\oi} \ / \cii \, (10$^{-2}$)       & $ 62.9 \pm 18.5$ & $ 81.5 \pm 18.2$ & $ 45.2 \pm 5.8$ & $ 37.1 \pm 11.7$ & $ 70.7 \pm 11.7$ \\ 
\textbf{\nii} \ / \cii \, (10$^{-2}$)       & [ 34.9 ] & $ 11.2 \pm 3.5$ & ( 4.0 ) & ( 23.6 ) & $ 7.4 \pm 2.0$ \\ 
\textbf{\oiii} \ / \cii \, (10$^{-2}$)      & [ 133 ] & $ 62 \pm 21$ & \nodata         & $ 56 \pm 11$ & $ 69 \pm 15$ \\ 
\textbf{\oiii} \ / \oi \, (10$^{-2}$)      & [ 108 ] & $ 118 \pm 47$ & \nodata         & $ 192 \pm 59$ & $ 176 \pm 62$ \\ 
\oiii \ / \textbf{\nii} \, (10$^{-2}$)      & ( 812 ) & ( 673 ) & \nodata         & -- & ( 825 ) \\
\\ \multicolumn{6}{c}{TIR} \\  
\hline \\
f$_{\mathrm{TIR},Spitzer}$ (10$^{-15}$ W m$^{-2}$)                    & $4.2 \pm 0.8$ & $15.4 \pm 0.8$ & $24.0 \pm 0.6$ & $7.9 \pm 1.4$ & $18.0 \pm 1.0$ \\ 
\\ 
\cii \ / TIR$_{Spitzer}$ (10$^{-2}$)       & $ 0.63 \pm 0.17$ & $ 0.37 \pm 0.06$ & $ 0.59 \pm 0.09$ & $ 0.86 \pm 0.20$ & $ 0.51 \pm 0.09$ \\ 
\oi \ / TIR$_{Spitzer}$ (10$^{-2}$)       & $ 0.10 \pm 0.04$ & $ 0.16 \pm 0.05$ & $ 0.22 \pm 0.04$ & $ 0.15 \pm 0.06$ & $ 0.19 \pm 0.04$ \\ 
\nii \ / TIR$_{Spitzer}$ (10$^{-2}$)       & $ 0.028 \pm 0.010$ & $ 0.016 \pm 0.005$ & ( 0.004 ) & ( 0.009 ) & $ 0.011 \pm 0.004$ \\ 
(\oi \ + \cii ) / TIR$_{Spitzer}$ (10$^{-2}$) & $ 0.7 \pm 0.2$ & $ 0.5 \pm 0.1$ & $ 0.8 \pm 0.1$ & $ 1.0 \pm 0.2$ & $ 0.7 \pm 0.1$ \\ 
\\ 
f$_{\mathrm{TIR},Herschel,100\mu m}$ (10$^{-15}$ W m$^{-2}$)                    & $2.2 \pm 0.4$ & $7.7 \pm 0.5$ & $14.8 \pm 0.7$ & \nodata         & \nodata         \\ 
\\ 
\cii \ / TIR$_{Herschel,100\mu m}$ (10$^{-2}$) & $ 1.19 \pm 0.31$ & $ 0.73 \pm 0.11$ & $ 0.95 \pm 0.13$ & \nodata         & \nodata         \\ 
\oi \ / TIR$_{Herschel,100\mu m}$ (10$^{-2}$) & $ 0.19 \pm 0.07$ & $ 0.33 \pm 0.10$ & $ 0.35 \pm 0.06$ & \nodata         & \nodata         \\ 
\nii \ / TIR$_{Herschel,100\mu m}$ (10$^{-2}$) & $ 0.054 \pm 0.019$ & $ 0.032 \pm 0.011$ & ( 0.007 ) & \nodata         & \nodata         \\ 
(\oi \ + \cii ) / TIR$_{Herschel,100\mu m}$ (10$^{-2}$) & $ 1.4 \pm 0.3$ & $ 1.1 \pm 0.2$ & $ 1.3 \pm 0.2$ & \nodata         & \nodata         \\ \\ 

f$_{60\mu m}$ / f$_{100\mu m}$             & $0.3 \pm 0.1$ & $0.6 \pm 0.1$ & $0.4 \pm 0.1$ & $0.1 \pm 0.3$ & $0.3 \pm 0.1$ \\ 
L$_{\mathrm{TIR},Spitzer}$ / L$_\mathrm{B}$             & $0.034 \pm 0.007$ & $0.022 \pm 0.002$ & $0.187 \pm 0.016$ & $0.035 \pm 0.006$ & $0.040 \pm 0.004$ \\ 
L$_{\mathrm{TIR},Herschel,100\mu m}$ / L$_\mathrm{B}$             & $0.017 \pm 0.003$ & $0.010 \pm 0.001$ & $0.107 \pm 0.006$ & \nodata         & \nodata         \\ 
L$_\mathrm{TIR,IRAS}$ / L$_\mathrm{B}$             & $0.136 \pm 0.036$ & $0.050 \pm 0.005$ & $0.325 \pm 0.047$ & $0.091 \pm 0.165$ & $0.151 \pm 0.033$ \\ 

\\

\enddata

\tablecomments{Upper limits (3$\sigma$) are given in parentheses, lower limits are given in square brackets. 
When comparing \oi, \oiii, and \nii\ data with \cii, the line data are first convolved to the beam size of the \cii\ map. These fluxes are noted with an asterisk (*).  The fluxes used in line ratios are taken from apertures matched to those of the species in boldface.  See \S\ref{sec:stackspec} and \S\ref{sec:TIRvsFIR} for a description of emission region selection and flux integration.
 } 
\end{deluxetable*}

Table ~\ref{table:lineratios} lists all of the image-total line fluxes and line ratios determined for each of our galaxies. The mean and dispersion for each flux ratio is given in Table~\ref{table:ratiomeandisp}.  All flux values were computed from maps that were first convolved to the limiting beam size for each ratio.  Since the ratios are simply computed from the total flux recovered over each footprint, they are unresolved and suited for comparison with the measurements reported by \citetalias{Brauher}.  Figure~\ref{fig:lineratios_internal} shows ratios of the various \herschel\ lines as a function of metallicity and the common FIR diagnostics $\mathrm{f_{60\mu m}/f_{100\mu m}}$ and $\mathrm{L_{TIR}/L_B}$ for our sample.  Unresolved 60$\mu$m/100$\mu$m ratios were often used in \iras\ studies, so we also use them to compare with the results from \citetalias{Brauher}.  It is important to note, though, that their values are for unresolved total galaxy fluxes, while we are looking at (usually small) regions within each of our galaxies.  Figure~\ref{fig:lineratios_tir} shows ratios of \herschel\ lines to TIR fluxes, compared with $\mathrm{f_{60\mu m}/f_{100\mu m}}$, $\mathrm{L_{TIR}/L_B}$, and metallicity.

{
\begin{figure*}[h!]
\centering
\includegraphics[width=\textwidth]{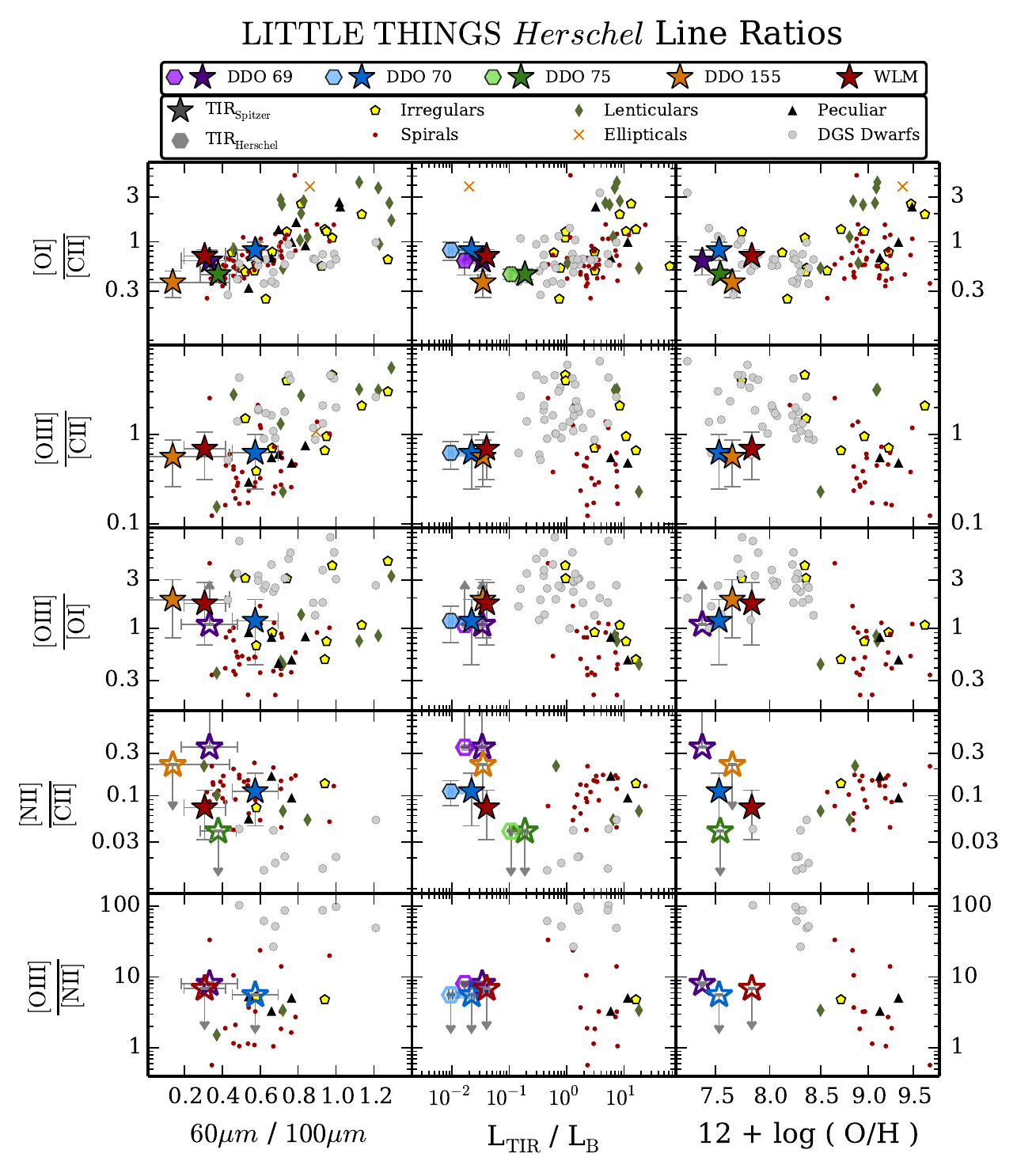}
\caption{\herschel\ far infrared line ratios.  Line fluxes were integrated over the regions shown in  Figure~\ref{fig:SEstackregs}.  The fluxes for each ratio were integrated over identical apertures, according to the selection order: \nii, \oiii, \oi. IRAS data for additional \lt\ galaxies come from \cite{HunterIRAS} and ISO data from \cite{HunterISO}.  Spiral, elliptical, and irregular$^\dagger$ galaxy data points from \cite{Brauher} are given for comparison with the $\mathrm{60\mu m/100\mu m}$ ratios, TIR/B ratios, and metallicities, complemented by dwarfs from the DGS sample \citep{Cormier2015}.   We only plot detections for the comparison sample data, omitting upper limits from \citetalias{Brauher}.   TIR for that sample are computed from the 25, 60, and 100 $\mu$m flux densities: TIR$_{IRAS} = 2.403(\nu f_\nu)_{25} -0.245(\nu f_\nu)_{60}+1.638(\nu f_\nu)_{100}$.   Metallicities for the reference sample were determined from the Luminosity-Metallicity relation described by \citet{Lamareille2004}.  For our measurements, detections are denoted by solid filled stars.  TIR derived from \herschel\ photometry is also shown by the hexagon symbols for galaxies with PACS 100$\mu$m data: DDO 69, DDO 70, and DDO 75. Upper and lower limits (3$\sigma$) for our data appear as empty symbols with arrows. $\\ \dagger$ The \citetalias{Brauher} sample irregulars consist almost entirely of dIrrs, many of which belong to the larger \lt\ sample.  They do include a few non-dwarf irregulars, however, such as Arp 220 and the Cartwheel Galaxy.}
\label{fig:lineratios_internal}
\end{figure*}

\begin{figure*}[]
\centering
\includegraphics[width=\textwidth]{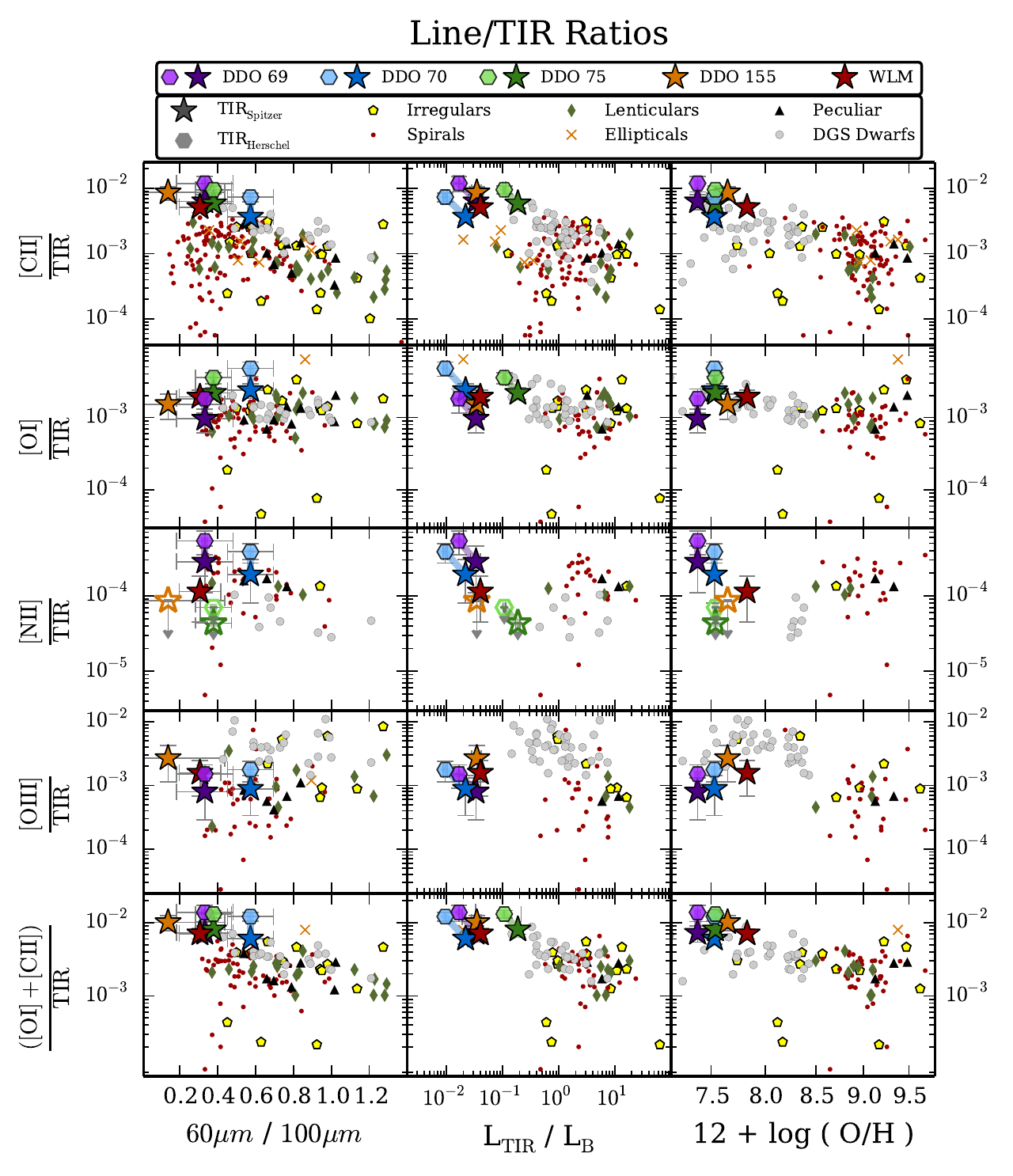}
\caption{PACS and Spitzer TIR line ratios. Line fluxes were integrated over the regions shown in  Figure~\ref{fig:SEstackregs}: the majority of the footprints for \cii\ and \oi, and smaller manually-identified regions for \oiii\ and \nii.  The \nii\ and \oiii\ error bars each have an additional 50\% uncertainty added in quadrature, to account for potential biases from the aperture selection.  The reference sample dwarf, spiral, elliptical, and irregular galaxy markers are the same as those described in Figure~\ref{fig:lineratios_internal}.  Again, we only plot those data points from \cite{Brauher} with published detections, and TIR for that sample is calculated from 25, 60, and 100 $\mu$m data.  Our detections, using TIR calculated from \spitzer\ maps, are represented by solid stars. The hexagons represent TIR determined from PACS 100$\mu$m, where available, and colored lines linking to the \spitzer\ points (stars) emphasize the differences.  Upper limits (3$\sigma$) appear as empty symbols.  }
\label{fig:lineratios_tir}
\end{figure*}
}

The 60$\mu$m/100$\mu$m ratio is an oft-cited diagnostic of the dust-heating intensity in a system -- to first order, the ratio depends on grain temperature.  This FIR color is therefore a proxy for dust temperature, and is typically associated with global star formation activity \citep{Dale01}.  The ratio of TIR and B luminosities represents the amount of light reprocessed by dust in the system compared to the starlight that escapes.  It is typically taken as an indicator of both star formation activity and optical extinction.  The interpretation of trends with TIR/B is admittedly complicated, as several factors can contribute to changes in this ratio.  It is still a useful endeavor to compare results with previous work in the literature, though, and we comment on different interpretations when appropriate.

When comparing the \herschel\ line fluxes with infrared emission from \spitzer, we estimate total uncertainties by adding the internal uncertainties within each image (\S~\ref{sec:uncertainties}) and calibration uncertainties from each instrument (Table~\ref{table:instrumentparams}) in quadrature.

\subsection{Spatial Comparison of FIR Line Emission}
\label{sec:regioncomp}

We generally find that regions of \oi\ and \oiii\ emission are closely located to regions of \cii\ emission, though they are not always precisely overlapping.  Refer to Figure~\ref{fig:coverage} for the flux maps of all five galaxies.   In targets where we detect \nii, the emission peaks fall right at the edge of the main \cii\ emission.  

DDO 69 has a bright \cii\ peak in the south of the field of view, with emission extending to the south and west.  The weak \oi\ peak is roughly cospatial with the western portion of the \cii, and weak \oiii\ is detected 30\arcsec \ to the northwest.  DDO 70 peaks in \cii\ in the center of the image, and is extended NE-SW.  The \oi\ emission peaks in the same region as the \cii, while the \oiii\ has one slight enhancement in the same position and another stronger one 30\arcsec \ to the north.  DDO 75 has two bright regions visible in the \cii\ map -- one in the middle and a smaller one to the SW, offset by $\sim$15\arcsec.  In each case, the \oi\ peak is cospatial with the \cii, but the central \oi\ knot is relatively thin in the N-S extent ($\sim$10\arcsec), while maintaining the $\sim$25\arcsec \ E-W extent of the \cii.  DDO 155 has two distinct \cii\ emission regions: one just east of the image center, and the other roughly 30\arcsec\ to the southwest.  The \oiii\ emission originates in similar locations, though slightly offset.  The \oi\ again appears somewhat similar, but the southwest component is not detected, and the \nii\ signal comes from just north of the SW component of \cii.   WLM shows a large, wide N-S column of \cii\ spanning the entire map.  The \oi\ peak lies just to the NW of the \cii\ and the \oiii\ just to the NE, near the north edge of the footprint. The \nii\ detection mainly comes from a region west of the center of the bulk of the \cii, but has a low absolute flux.

\subsection{Comparison with Multi-Wavelength Data}
\label{sec:obscomp}

\begin{figure*}
\centering
\includegraphics[width=\textwidth]{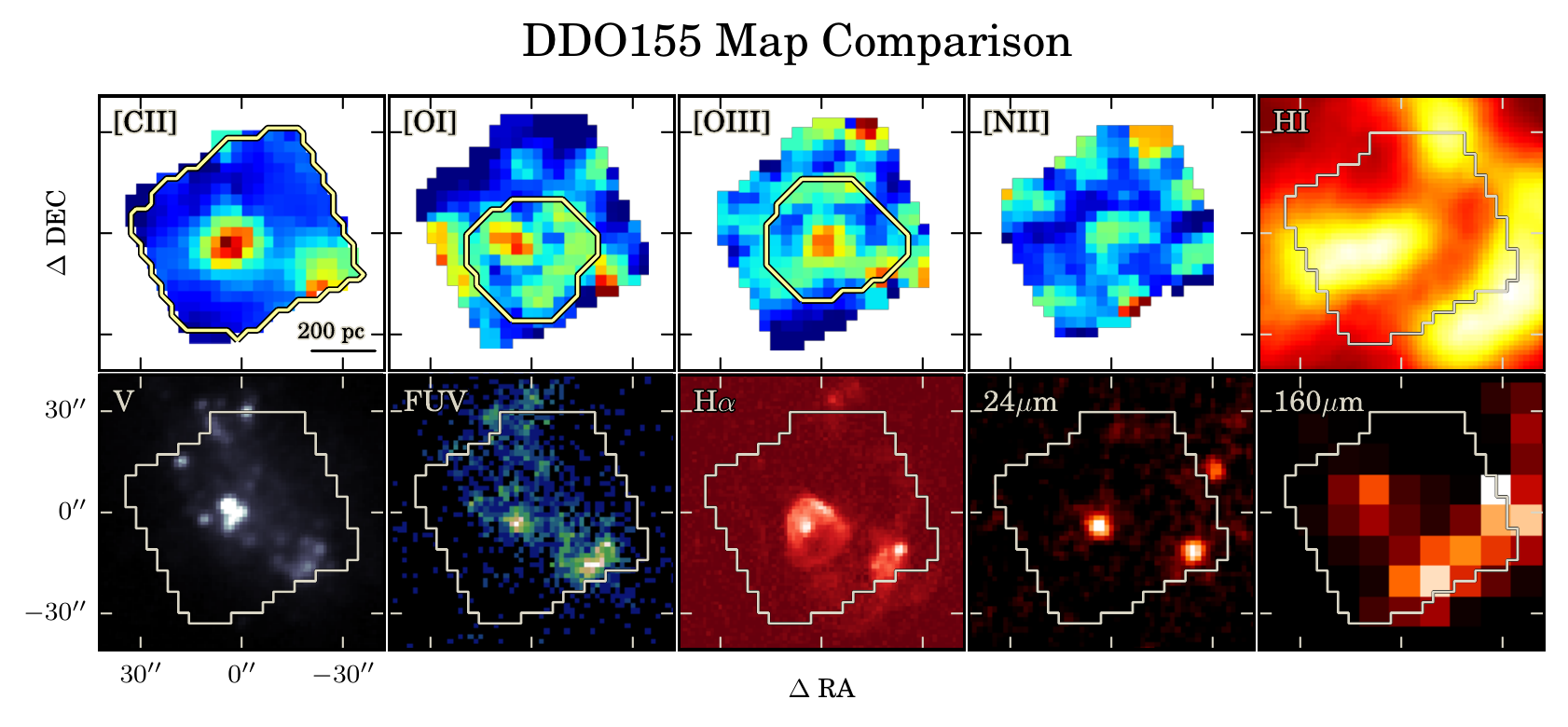}
\caption{Spatial comparison of emission in the \herschel\ PACSman maps with other observations at several wavelengths.  DDO 155 appears here, with the other galaxies presented in Figures~\ref{fig:mapcomp2} and~\ref{fig:mapcomp3}. Each image scaling is independent, to enhance dynamic range and draw out faint features.  Brightness (yellow) in the \hi\ colormap now corresponds to higher column density (as opposed to darkness in the grayscale of Figures~\ref{fig:zoom} and~\ref{fig:zoomonline}). The \herschel\ maps have overlaid contours showing regions where S/N$>$3 for the flux contained within regions spanning a minimum of 2 beam widths, as described in \S~\ref{sec:stackspec}. Since our signal is typically weak (except for \cii), and since we are interested in detections aggregated from \textit{regions} the size of one beam or larger, we do not look at S/N values of single re-gridded pixels. In the other maps, the white outlines indicate the footprint of the \cii\ map.  Here, \hi\ is from \cite{LTdata}, $V-$band from \citetalias{HunterUBV}, FUV from \cite{Dale09}, H$\alpha$ from \citetalias{HunterHa}, 24$\mu$m \& 160$\mu$m from \cite{Dale09}. }
\label{fig:mapcomp}
\end{figure*}

\begin{figure*}
\centering
	\begin{subfigure}[]{}
		\includegraphics[width=0.95\textwidth]{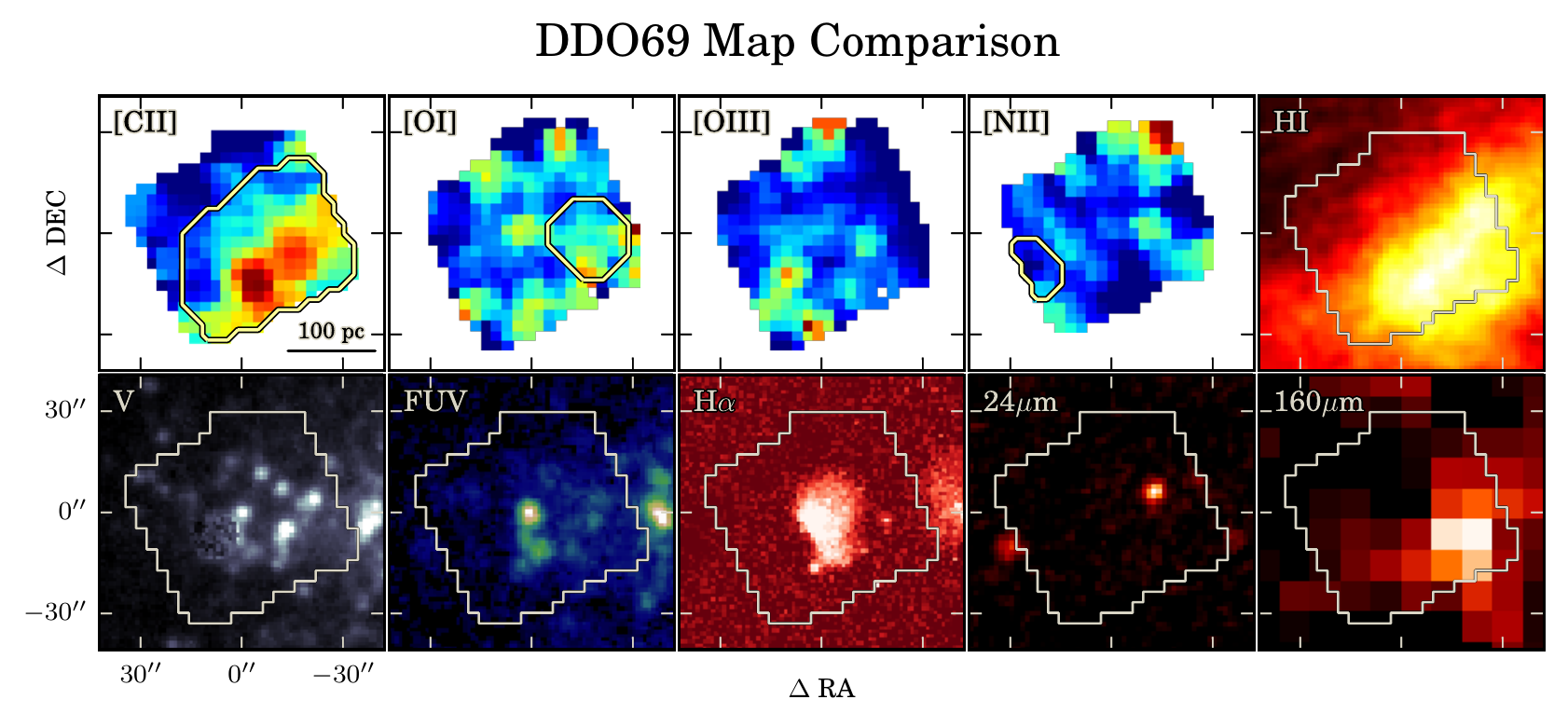}
	\end{subfigure}	
	\\
	\begin{subfigure}[]{}
		\includegraphics[width=0.95\textwidth]{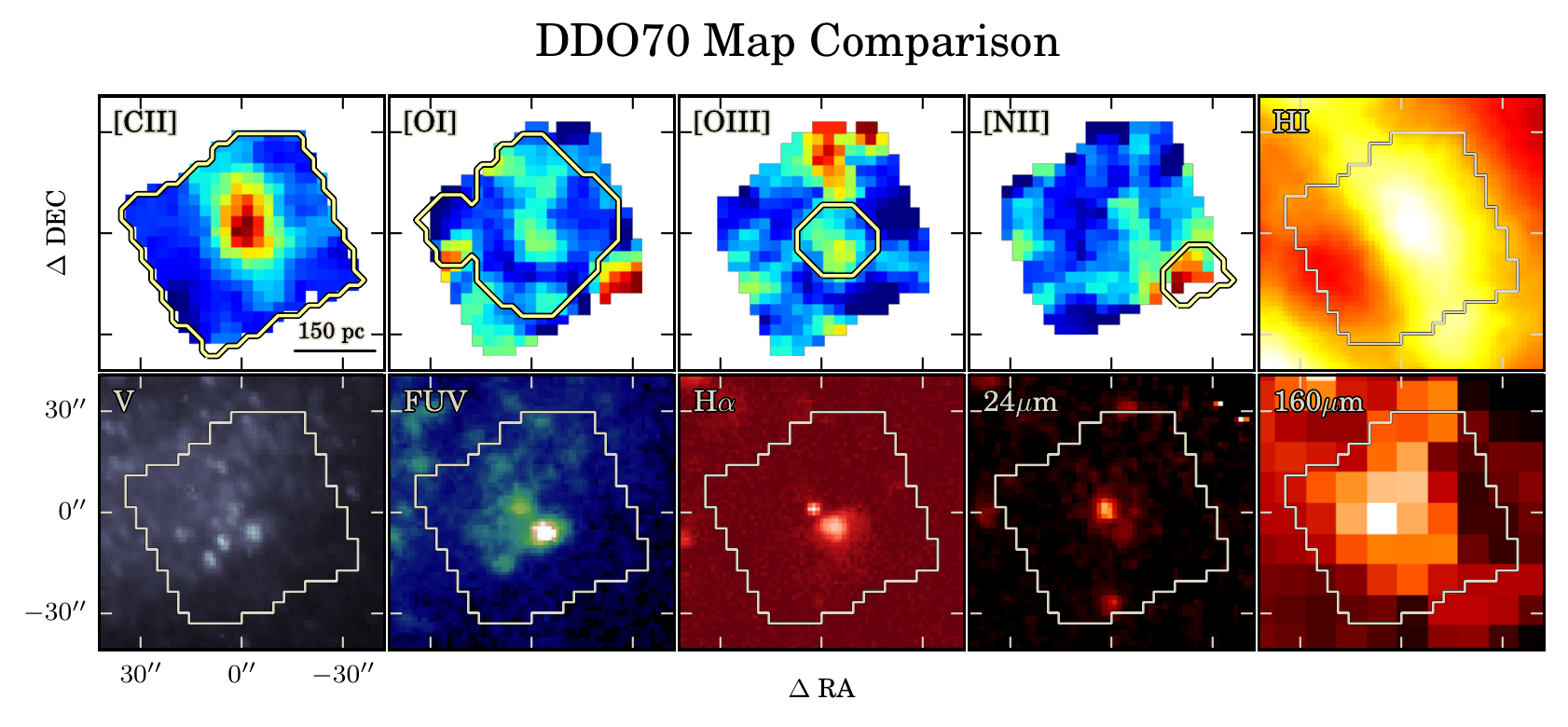}
	\end{subfigure}	
\caption{Comparison of emission in the \herschel\ PACSman maps with other observations at several wavelengths.  
\hi\ is from \cite{LTdata}, $V-$band from \citetalias{HunterUBV}, FUV from \cite{HunterFUV}, H$\alpha$ from \citetalias{HunterHa}, 24$\mu$m \& 160$\mu$m from \cite{Dale09}. We note that, in general, \cii\ emission correlates well with \hi\ emission, and bright knots in the bands that trace star formation are often near \cii\ peaks. 
}
\label{fig:mapcomp2}
\end{figure*}

\begin{figure*}
\centering
	\begin{subfigure}[]{}
		\includegraphics[width=0.95\textwidth]{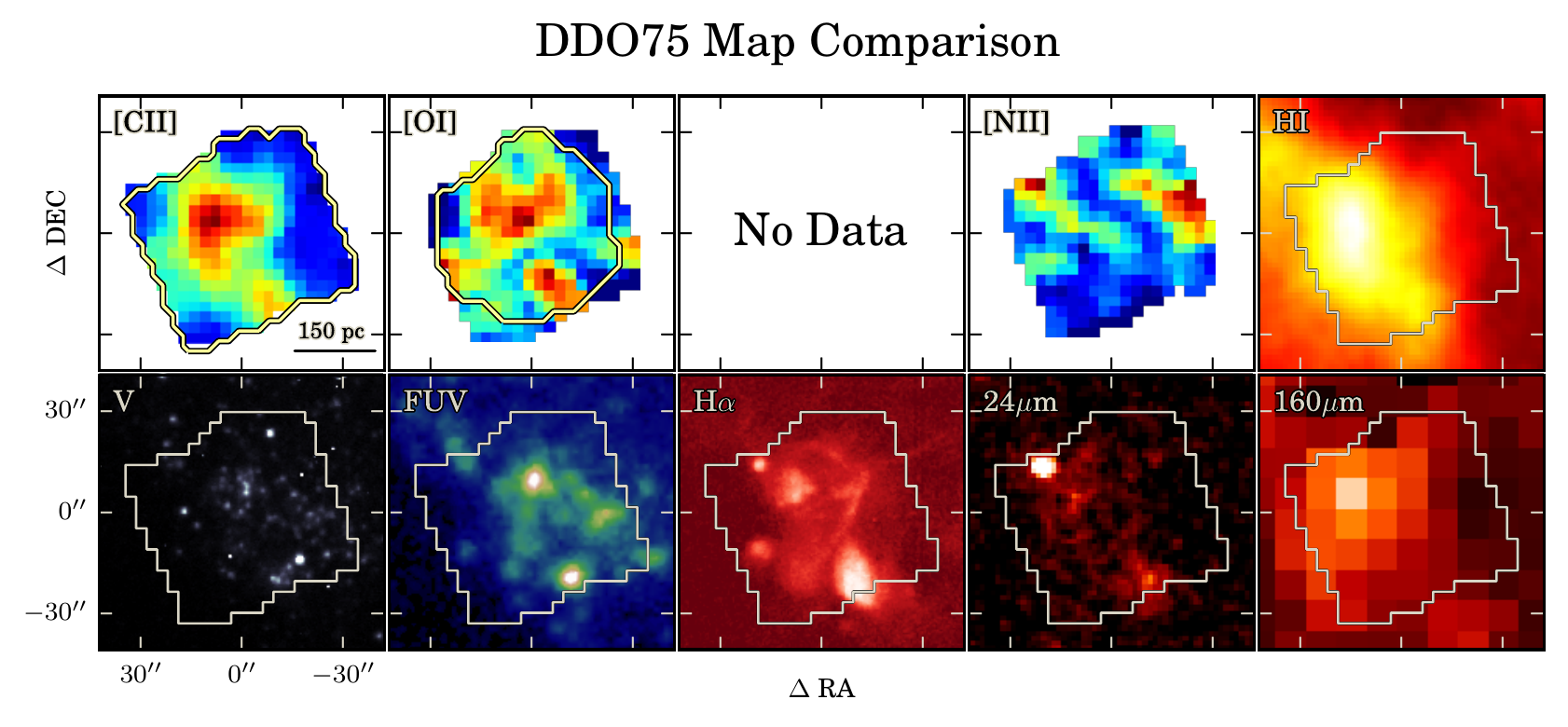}
	\end{subfigure}	
	\\
	\begin{subfigure}[]{}
		\includegraphics[width=0.95\textwidth]{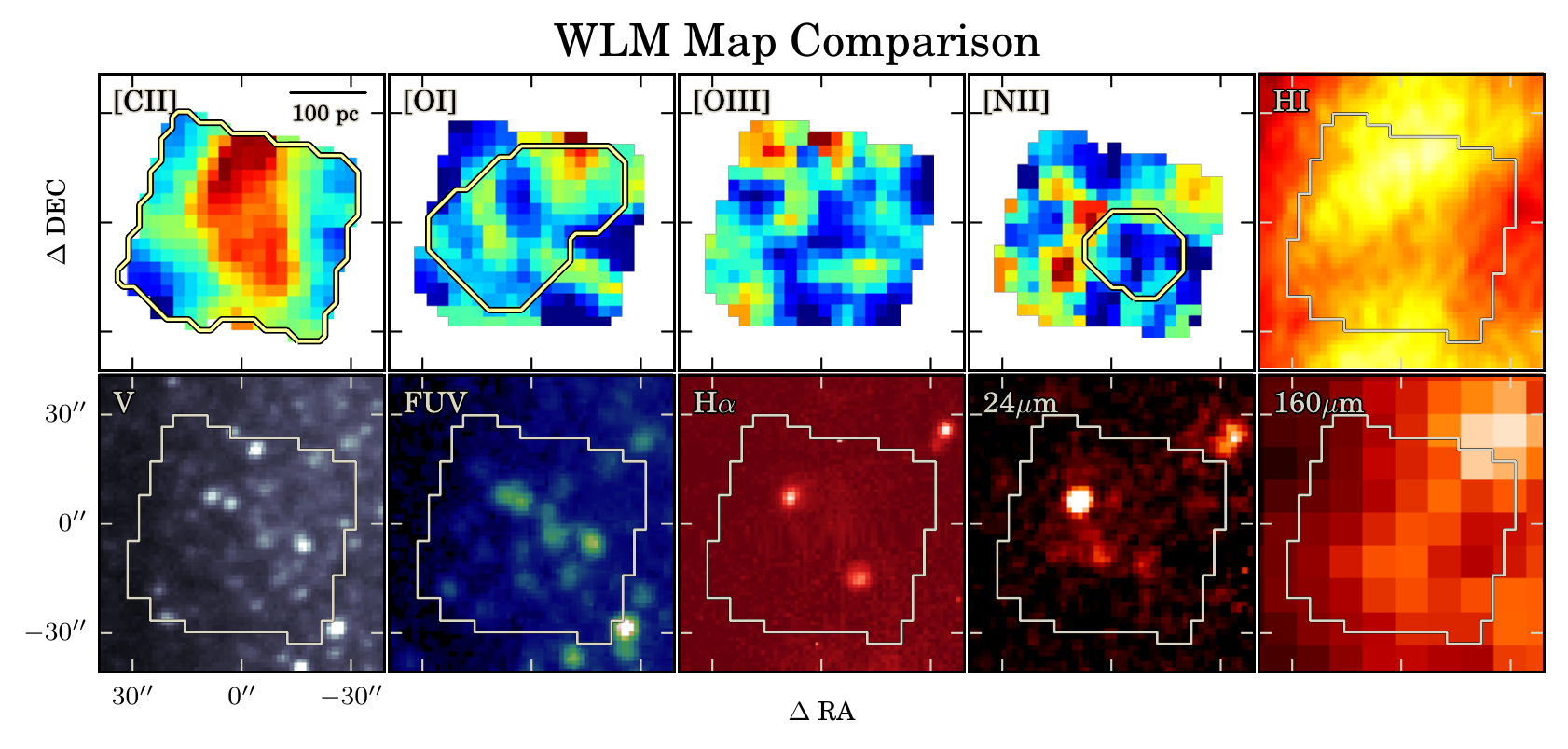}
	\end{subfigure}
\caption{Comparison of emission in the \herschel\ maps with other observations at several wavelengths.  \hi\ is from \cite{LTdata}, $V-$band from \citetalias{HunterUBV}, FUV from \cite{HunterFUV}, H$\alpha$ from \citetalias{HunterHa}, 24$\mu$m \& 160$\mu$m from \cite{Dale09}. We note that, in general, \cii\ emission correlates well with \hi\ emission, and bright knots in the bands that trace star formation are often near \cii\ peaks. 
}
\label{fig:mapcomp3}
\end{figure*}

\begin{figure}
\centering
\includegraphics[width=.5\textwidth]{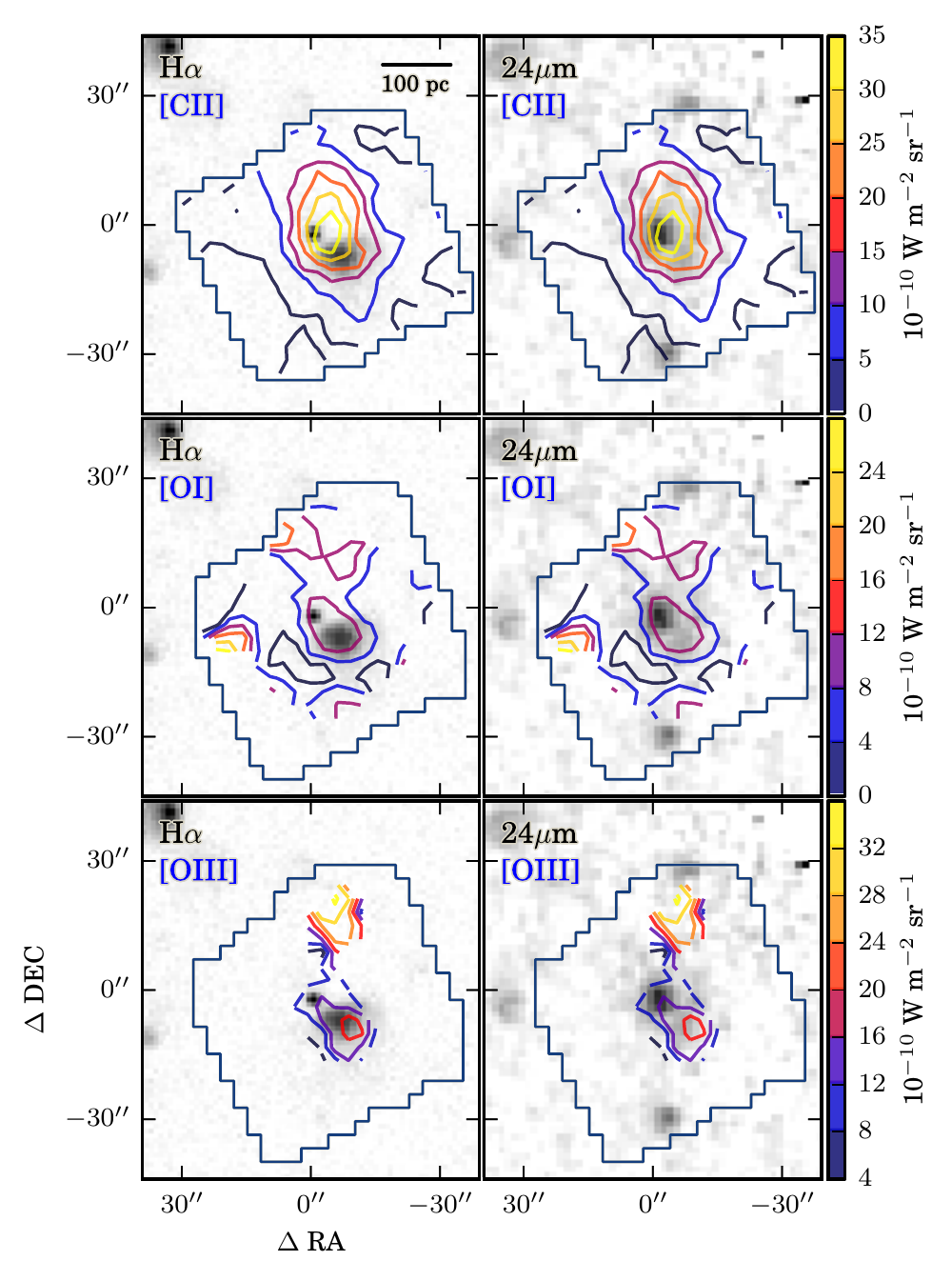}
\caption{ \cii, \oi, and \oiii\ intensity maps overlaid onto H$\alpha$ and 24$\mu$m grayscales (dark = brighter emission) for DDO 70. The \herschel\ contours are the same in each row.  In this galaxy, the \cii\ emission is offset from the H$\alpha$ peaks, while the \oiii\ emission is more closely aligned to H$\alpha$.  It is interesting to note that the H$\alpha$ and 24$\mu$m peaks are slightly offset.   }
\label{fig:ddo70overlays}
\end{figure}

\begin{figure*}
\centering
\includegraphics[width=\textwidth]{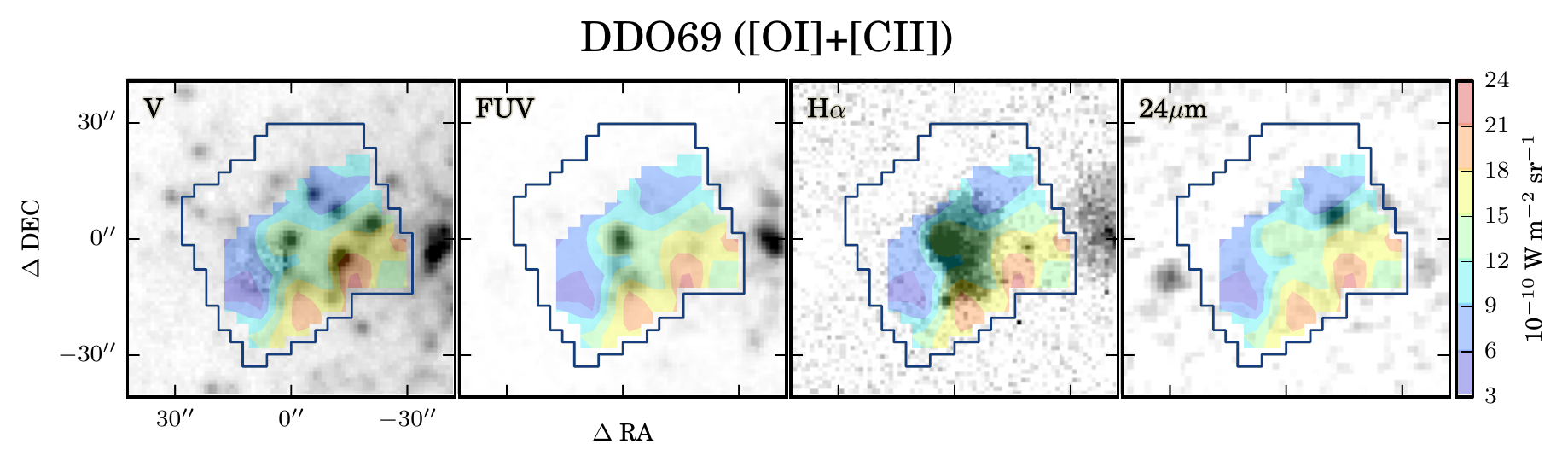}
\includegraphics[width=\textwidth]{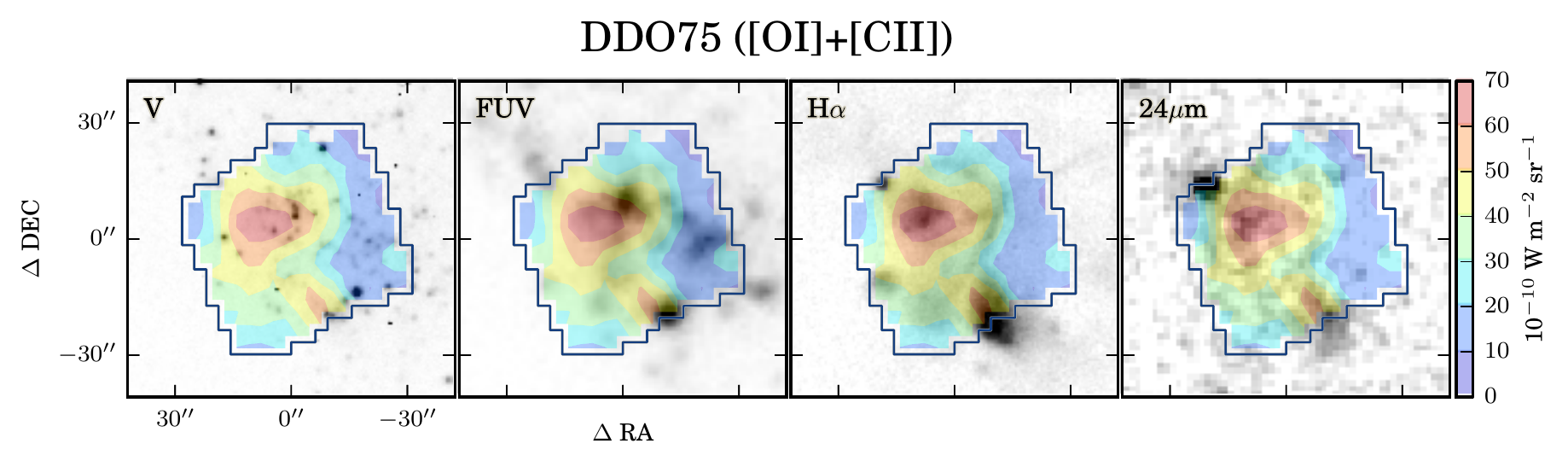}
\caption{ (\oi + \cii) intensity overlaid onto V, FUV, H$\alpha$, and 24$\mu$m maps in grayscale (dark = brighter emission) for DDO 69 and DDO 75. The filled \herschel\ contours are the same across each row.  In DDO 69, the (\oi + \cii) emission is concentrated at the edge of the H$\alpha$ region - a nice example of a PDR.  Agreement is good for H$\alpha$ in DDO 75, but slight offsets are apparent for FUV.  }
\label{fig:oiplusciioverlays}
\end{figure*}

There are some qualitative trends apparent in the comparison of maps at different wavelengths of the same regions in each galaxy.  To visualize this, we plot our \herschel\ maps side-by-side with \hi, $V-$band, FUV, H$\alpha$, and \spitzer\ MIPS 24$\mu$m and 160$\mu$m images (see Figure~\ref{fig:mapcomp} for DDO 155 and Figures~\ref{fig:mapcomp2} \&~\ref{fig:mapcomp3} for the rest).  For DDO 155, the two bright regions that appear in the \cii\ map are clearly identifiable in all but the \nii\ and 160$\mu$m images.   Some 3-5$\sigma$ features in the fainter PACS maps do not have corresponding emission that may be expected in other bands.

One notable feature in many of our maps is that the \cii\ peaks are often near, but not completely aligned with, the FUV, H$\alpha$, and 24$\mu$m knots, all indicators of current star formation.  The \cii\ is generally more closely aligned with the 24$\mu$m emission than with H$\alpha$.  The \oiii\ peaks, on the other hand, are usually on top of the H$\alpha$ peaks.  

The overlays in Figure~\ref{fig:ddo70overlays} show this clearly for DDO 70. While the H$\alpha$ features are cospatial with some \cii\ emission, the \cii\ peak is offset between the two H$\alpha$ features and is better aligned with the 24$\mu$m emission.  \oiii\ peaks more strongly on the more extended southern \hii\ region than on the compact \hii\ region coincident with the 24$\mu$m source. These maps therefore present an excellent example of a PDR where the highly ionized gas (\oiii) is in the south and the neutral gas (24$\mu$m and \cii) is in the north.  For reference, the southern \hii\ region is roughly 70 pc across with an average H$\alpha$ brightness of 2.55$\times10^{-9}$ W m$^{-2}$ sr$^{-1}$ -- almost identical in brightness to \oiii\ in that region.  The more compact northern \hii\ region is about 30 pc across, with an average H$\alpha$ brightness of 2.93$\times10^{-9}$ W m$^{-2}$ sr$^{-1}$, similar to that of the cospatial \cii .

Combining \cii\ and \oi\ gives a measure of the total FIR cooling.  Since \cii\ is the dominant coolant in our galaxies, (\oi + \cii) closely resembles \cii.  The peaks of the (\oi + \cii) maps are often cospatial with peaks of some star formation tracers but slightly offset from other tracers within the same system.  For example, the two (\oi + \cii) maxima in DDO 75 coincide with the local H$\alpha$ maxima, and to the same but fainter regions in the 24$\mu$m emission, but are offset by 5-10\arcsec\ from the bright knots in the FUV.  In DDO 69, however, the (\oi + \cii) nicely traces the outer edge of the bright H$\alpha$ region, indicating rough PDR boundaries.  This is shown in Figure~\ref{fig:oiplusciioverlays}. 

Of further interest is the fact that the \cii\ peaks are cospatial with regions of high \hi\ column density (Figures~\ref{fig:mapcomp} --~\ref{fig:mapcomp3}).  The \cii-\hi\ correlation is striking in the rest of our sample, as well. DDO 69, in particular, exhibits an almost perfect spatial agreement between the two.  DDO 70, DDO 75, and WLM show the \hi\ offset by only 5-10\arcsec\ from the \cii\ peaks with the broader emission overlapping closely.  The star formation tracers are offset as well, but nearby (within $\sim$15\arcsec) for every observed region.  We discuss the 2D relationship with \hi\ in more detail in our forthcoming paper (Cigan et al., in preparation). 
  
Recent work by \cite{DeLooze2014} has demonstrated that several fine structure lines accessible to PACS correlate with FUV+24$\mu$m star formation.  Our galaxies are consistent with their findings.  We calculated the extinction-corrected $\Sigma_\mathrm{SFR}$ based on a weighted combination of FUV and 24$\mu$m using their prescription \citep[originally from][]{Hao11} as well as the prescription of \cite{Leroy12}, for comparison.  For the image-total \cii\ surface densities in our systems, the \cite{DeLooze2014} formula predicts an average log $\Sigma_\mathrm{SFR(FUV+24)}$ of -2.75 M$_\odot$ yr$^{-1}$ kpc$^{-2}$ for our galaxies.  The observed values (\mbox{-2.55} using the calibration of \citet{Leroy12} or \mbox{-2.44} using that of \citet{Hao11}) are both within the 0.32 dex scatter reported by \cite{DeLooze2014}.

\section{DISCUSSION}
\label{sec:discussion}

\subsection{Origins of FIR Line Emission}
\label{sec:firlines}

The \cii\ 158$\mu$m line is one of the primary FIR coolants of the ISM, typically accounting for about 0.1-1\% of the FIR energy budget \citep[e.g.,][]{Stacey91}.  Prevalent in radiation fields with photon energies between 11.26-24.38 eV, \cii\ can be found in diffuse ionized regions as well as denser regions with molecular gas.  Specifically, \cii\ can exist at the surface of molecular clouds, which can also harbor dust- or self-shielded \hi\ and H$_2$ (\citealp{TH85}, \citealp{Wolfire10}).  We explore the relation between carbon and hydrogen species in our galaxies in an upcoming paper.  See \cite{HT97} for their classic review of PDR structure and atomic/molecular species arrangement therein.  Also see \cite{Wolfire90}, \cite{Bolatto99}, and \cite{Kaufman99} for additional discussions of the models of FIR line tracers of PDRs, and \cite{Rollig06} who address PDRs at low metallicities.  

\oi\ emission at 63$\mu$m originates in the warm and cold neutral media.  Neutral oxygen's ionization potential of 13.62 eV is quite close to that of hydrogen, making it a good tracer of the boundaries of neutral regions.  This line is one of the main coolants of warm, dense gas, and traces gas deeper into PDRs than \cii\ because it has a higher critical density for collisions with hydrogen.  

The \oiii\ 88$\mu$m line comes from highly ionized environments, with radiation energies $>$35.12 eV.  This comes from early O stars in \hii\ regions.  

\nii\ 122$\mu$m emission comes from diffuse, ionized regions, where the radiation is more energetic than the 14.53 eV required to ionize neutral nitrogen. The critical density of \nii\ is low ($n\sim 300$), so this does not arise in denser \hii\ regions.  Since some of the \cii\ observed can come from this phase, \nii\ can be used to separate the diffuse \cii\ component from the PDR emission \citep{Heiles94}.

\subsection{Whole Image Trends}
\label{sec:largetrends}

\begin{deluxetable}{ lcccc }
\tablecaption{ Line Ratio Statistics \label{table:ratiomeandisp}} 
\tablehead{ \colhead{Ratio} & \colhead{Min.} & \colhead{Max.} & \colhead{Mean} & \colhead{Standard Dev.}  }

\startdata
\oi/\cii       & 0.37 & 0.81 & 0.59  & 0.16  \\ 
\oiii/\cii     & 0.56 & 0.69 & 0.62  & 0.05  \\ 
\oiii/\oi      & 1.18 & 1.92 & 1.62  & 0.32  \\ 
\nii/\cii     & 0.07 & 0.11 & 0.093  & 0.019  \\ 
\\ 
  \multicolumn{5}{c}{TIR from MIPS} 
  \\  \\  
\cii/TIR       & 0.0037 & 0.0086 & 0.0059  & 0.0016  \\ 
\oi/TIR        & 0.0010 & 0.0022 & 0.0016  & 0.0004  \\ 
\oiii/TIR      & 0.0008 & 0.0027 & 0.0015  & 0.0007  \\ 
\nii/TIR      & 0.00011 & 0.00028 & 0.00019  & 0.00007  \\ 
(\oi+\cii)/TIR & 0.0053 & 0.0101 & 0.0076  & 0.0016  \\ 
\\ 
  \multicolumn{5}{c}{TIR from PACS} 
  \\  \\ 
\cii/TIR       & 0.0073 & 0.0119 & 0.0096  & 0.0019  \\ 
\oi/TIR        & 0.0036 & 0.0048 & 0.0042  & 0.0006  \\ 
\oiii/TIR      & 0.0022 & 0.0150 & 0.0086  & 0.0064  \\ 
\nii/TIR      & 0.00038 & 0.00203 & 0.00120  & 0.00083  \\ 
(\oi+\cii)/TIR & 0.0121 & 0.0138 & 0.0130  & 0.0007  \\ 
\enddata
\tablecomments{Minimum, maximum, mean and standard deviation values. PACS line ratios and PACS line -- PACS TIR ratios were determined from matched apertures.  \spitzer\ TIR was determined for the whole PACS field of view. Only detections are included in the calculations.  }
\end{deluxetable}

Several previous studies have probed the relations of the various FIR lines to FIR colors in dwarfs, such as \citet[hereafter \citetalias{Mal01}]{Mal01}, \citetalias{HunterISO}, and \citetalias{Brauher}. These previous works were based on spatially unresolved observations.  The large beam sizes of \kao\ (FWHM $\sim$55\arcsec), \iso\ (FWHM $\sim$75\arcsec), and \iras\ (FWHM $\sim$90\arcsec) in these earlier studies meant that a single measurement could cover the majority or entirety of a dwarf galaxy.  Our single pointings represent small regions (200-500pc on each side) within a particular galaxy, and while each footprint can include several large star-forming complexes, we are still probing much smaller scales than the pre-\herschel\ works. Newer studies utilizing \herschel, such as DGS \citep[][hereafter \citetalias{Cormier2015}]{Cormier2015} and KINGFISH \citep{KINGFISH}, often have a mix of single pointings and mosaicked maps bridging the gap between the small scales of individual PDRs mapped in Galactic studies and large galaxy-average scales that limited previous-generation instruments.  Our sample is the first to probe the FIR line emission of normal dwarfs with low metallicity and moderate star formation rates at these spatial scales.  

Figure~\ref{fig:lineratios_internal} shows various PACS FIR line ratios -- \oi/\cii, \oiii/\cii, \oiii/\oi, and  \nii/\cii\ plotted against 60$\mu$m/100$\mu$m, L$_\mathrm{TIR}$/L$_\mathrm{B}$, and metallicity. Figure~\ref{fig:lineratios_tir} shows ratios of PACS lines to integrated TIR fluxes from \spitzer\ -- \cii/TIR, \oi/TIR, \oiii/TIR, \nii/TIR, and (\oi+\cii)/TIR.  The line fluxes used in our ratios are not integrated over a whole galaxy or whole field of view, but are rather computed for smaller apertures matched between the lines as described in \S\ref{sec:stackspec}.  We include a comparison sample for many galaxy types originally presented by \citetalias{Brauher}.

The FIR color $\mathrm{60\mu m/100\mu m}$, often taken as a proxy for dust temperature or dust heating intensity, is typically low in our sample -- averaging around 0.3 -- falling on the low end of the \citetalias{Brauher} sample, though the errors in the IRAS observations are relatively large.   Low $60\mu m/100\mu m$ ratios (low dust temperature) in dwarfs have been discussed in the literature:  \citetalias{HunterISO}, in their comparison of \cii/f$_{15\mu \mathrm{m}}$ to $\mathrm{60\mu m/100\mu m}$, interpreted the dust as being cooler than in spirals because it's farther away, on average, from hot stars.  \cite{RemyRuyer13}, on the other hand, have noted in their recent dust models of DGS dwarfs and KINGFISH spirals that dwarfs have warmer median dust temperatures (32K, vs. 23K for spirals), and that the lowest-metallicity dwarfs have the highest dust temperatures.  This is supported by the findings of \cite{Walter07} that the low-Z, quiescent star-forming dIrrs in the M81 group have warmer dust temperatures than spirals.   These two results are not necessarily in disagreement -- the DGS sample has many dwarfs with lower metallicities than the \citetalias{HunterISO} sample.  Furthermore, selection effects of the targets of the different samples such as differences in $\Sigma_\mathrm{SFR}$ and the ionizing radiation fields can affect these ratios.

We also note quite low ratios of TIR/B in much of our sample -- so low that they appear to be truly set apart from the majority of systems discussed in previous works.  There are only a handful of galaxies from the \citetalias{Brauher} sample in this regime, including some ellipticals and fellow \lt\ dwarf IC 1613.  The average $B-$band and TIR luminosities in the \citetalias{Brauher} sample are roughly the same for each galaxy type, almost always within an order of magnitude.  Using the photometry parameters outlined in \citetalias{HunterUBV} to integrate over each entire galaxy, the average $B-$band luminosity of 2.4$\times10^{8}$ L$_\odot$ for our sample is two orders of magnitude lower than that of the \citetalias{Brauher} irregulars, though a handful of irregulars in their sample have similar L$_B$ values to ours.  Our galaxy-wide TIR luminosities, however, average roughly five orders of magnitude less than those of the \citetalias{Brauher} irregulars.  This indicates that our galaxies have \textit{low TIR}, rather than \textit{enhanced B}.

This could simply result from a reduction in thermal FIR radiation due to lower levels of dust and cool gas, or it could be caused by cooler dust temperatures.  The TIR continuum in our galaxies is low compared to the spectral line emission, but not nearly enough to account for the two orders of magnitude separating our TIR/B values from those in previous studies.  Since $B-$band emission is not particularly high, the low TIR relative to B suggests that there is little extinction, or less starlight is processed by dust, which is at least partially a result of the low metallicities -- and lower dust content -- of these galaxies. Our galaxies have much lower metallicity than the majority of the \citetalias{Brauher} sample, which supports this conclusion.  We will explore dust temperatures of our systems in detail in our forthcoming paper on the \herschel\ photometry of \lt\ galaxies.

There are no strong trends with metallicity among the detected line ratios within our sample.  Considering our sample in comparison to the \citetalias{Brauher} galaxies, there appear to be trends with \oi/\cii, \nii/\cii, and \cii/TIR. 

We now discuss the results of each set of line ratios in detail, with a synopsis at the end:

\textbf{\oi/\cii}:  \textit{We observe low to moderate \oi\ relative to \cii\ in our sources, suggesting cool environments.}  When compared to the $\mathrm{60\mu m/100\mu m}$ ratio, our galaxies appear to lay in a region more typical of spirals than dwarf irregular galaxies, as shown in Figure~\ref{fig:lineratios_internal}.  Though both ratios are low in our galaxies, there is a large uncertainty in the $\mathrm{60\mu m/100\mu m}$ ratio, which means the targets studied here could potentially be consistent with the low end of the other irregulars.  

We note that within our sample, \oi/\cii\ increases with dust heating intensity as traced by $\mathrm{60\mu m/100\mu m}$, and this fits in nicely with the trend visible among the other galaxies in the \citetalias{Brauher} and DGS samples.  \citetalias{Mal01} and \citetalias{Brauher} describe a strong positive correlation, with \oi\ dominating the ISM cooling in warmer environments (larger $\mathrm{60\mu m/100\mu m}$ ratio).  The transition point they find is $60\mu$m/$100\mu$m$\approx$0.8.  Our sample falls well below this mark, and thus our ratios of less than unity (\cii-dominant) are consistent with the trend.   We find no apparent trend within our galaxies for TIR/B, but our dwarfs fall more or less in line with the overall increasing trend combining other irregulars and spirals of the \citetalias{Brauher} sample.

The low metallicities in our systems may also be a key factor for cooler PDRs. There is less dust, which translates to less PDR gas heating, and thus lower average PDR temperatures.  According to the models of \cite{Rollig06}, this can result in more \cii\ and less \oi , independently of density.  Careful modeling will be required to determine the exact roles of temperature and density, however.   \oi/\cii\ appears to decrease slightly as metallicity goes down across several galaxy types, but the relation among irregulars has less scatter.  

\textbf{\oiii/\cii}: \textit{Our sources exhibit low ratios, among the lowest observed for dwarfs.} One striking result is the fact that we see low \oiii-to-\cii\ ratios (less than unity) compared with other dwarf galaxies, and pushing into a regime more typical of spirals.  Previous works (\citetalias{HunterISO}; \citealt{Cor10,Cor12,Leb12,Cormier2015}) have shown typical values of \oiii/\cii\ in dwarfs to be in the range of two to three times what we observe.  \oiii\ is often the brightest FIR line observed in low-metallicity starburst galaxies, but our galaxies are much more quiescent star-formers.  \oiii\ is produced by radiation from O and B stars in \hii\ regions; the low ratio could indicate a dearth of these massive stars near our observed regions, or in the systems in general. This is not clearly linked to decreasing metallicity -- there is no definitive trend in our data with Z.  Not only are the TIR/B values in our sample lower than for other galaxies, but the \oiii/\cii\ ratios are much lower than would be expected for the negative trend among dwarfs noted by \citetalias{Brauher}.

\textbf{\oiii/\oi}:  \textit{ Neutral oxygen dominates in most of our sources.}
Our sample shows a slight decreasing trend of \oiii/\oi\ with $60\mu m/100\mu m$, but they agree within the large errors. When viewed in the context of the \citetalias{Brauher} and DGS samples, they once again occupy an area between spirals (low oxygen ratio, low dust temperature) and irregulars (high ratio, high dust temperature), and there is no trend with galaxy type. \citetalias{HunterISO} and \citetalias{Brauher} noted that irregulars show an elevated \oiii/\oi \ ratio compared with spirals, but our data overlap with both.   Our galaxies do not follow the noted relation with TIR/B seen in the \citetalias{Brauher} sample of irregulars, and indeed appear to be completely separate from the other populations.

\textbf{\nii/\cii}:  \textit{This ratio favors PDRs as the primary source of \cii\ emission.  }  
Our dwarfs have similar ratios to the \citetalias{Brauher} detections, but somewhat higher ratios compared to the DGS dwarfs.   The \nii/\cii\ ratio can be quite sensitive to variations in \nii\ from local ionization levels.  But since \nii\ only comes from the diffuse ionized gas (and not PDR gas), the ratio of \nii\ to \cii\ can also decrease for environments where a smaller fraction of the \cii\ emission comes from \hii\ regions.  We might expect, therefore, to see a decreasing trend with decreasing heating intensity.  However, no such trend is clear in the \nii/\cii\ ratios for changing $60\mu m/100\mu m$ dust temperature, which is consistent with the findings of \citetalias{Mal01} and \citetalias{Brauher}. There is little scatter among our detections as TIR/B increases.  If we consider the upper limit for DDO 75, though, it appears that those in our sample with higher \nii/\cii\ could have lower TIR/B -- suggesting that as extinction increases, the fraction of diffuse ionized component of \cii\ emission decreases.  

It is possible that the majority of the nitrogen in these regions is in the neutral phase, with little ionizing radiation above 14.53 eV.  However, the H$\alpha$ maps show that there are \hii\ regions within each \herschel\ footprint, and we expect that any nitrogen in \hii\ regions should be in the ionized phase.  \cite{Cor12} recently published FIR line fluxes for Haro 11, finding \nii/\cii\ $\simeq$ 0.04, and the handful of galaxies in the DGS where \nii\ was detected have a median ratio of about 0.025 \citepalias{Cormier2015}.  The \iso\ observations of NGC 1156, NGC 1569, and IC 4662 discussed by \citetalias{HunterISO} provide only upper limits to \nii\ fluxes, giving similar \nii/\cii\ ratios between 2-8\% for those dwarfs.  The theoretical \nii/\cii\ ratios for densities $n_e \lesssim 1000$cm$^{-3}$ only go down to about 0.1 \citep[e.g.,][]{BernardSalas2012}, indicating that ionized gas is not the excitation source for \cii\ for ratios below this. All of the evidence from the \nii/\cii\ ratio supports the picture of \cii\ primarily originating in PDRs rather than the diffuse ionized medium.  This finding is important because it means that corrections to \cii/TIR for \cii\ emission from ionized gas are small, which has consequences for PDR modeling.  We will explore this topic further in an upcoming paper.

\textbf{\oiii/\nii}: \textit{Our limits have moderate ratios, suggesting softer radiation fields than in other dwarfs.}  Though we have detected both \oiii\ and \nii\ in our galaxies, neither is detected where the other is found, resulting in limits for the ratios.   Both species originate in ionized regions.  \oiii\ comes from regions of slightly higher density than \nii, however ($n_e\sim$500 and 300 cm$^{-3}$, respectively).  The different energies required to create them -- 14.5eV for \nii, 35.1eV for \oiii\ -- can also tell us about the hardness of the radiation field. The median value of 86.3 in the DGS sample \citep{Cormier2015}  is much higher than in any system of our sample, suggesting the radiation field is less hard.  \citetalias{Brauher} only hint at possible correlations with 60$\mu$m/100$\mu$m and TIR/B for their sample, due to insufficient detections.  Aside from the radiation field, variations in the overall elemental abundance ratios (N/O) can also affect this ratio.

\textbf{\cii/TIR}:  \textit{\cii\ can comprise around 1\% of the TIR emission in our dwarfs -- quite high relative to most other galaxies of any type. }  
With \cii\ levels around 0.5\% of TIR$_\spitzer$ for most of our sample, and near 1\% of the \herschel-based TIR, all of our galaxies except DDO 70 fall above the \citetalias{Brauher} sample -- the vast majority of other galaxies of all morphologies have \cii/TIR ratios of less than 0.4\%.  Considering that a typical high \cii-to-\textit{FIR} ratio for irregulars is $\approx$1\% \citep{Madden2000}, and that TIR is often about two times greater than FIR in dwarfs (\citetalias{HunterISO}), our galaxies have high ratios indeed. For context, the highest reported \cii/FIR detection of 2.2\% comes from the N159 region of the Large Magellanic Cloud \citep{Israel96}, using FIR determined from \iras .  \cite{DiazSantos14} find that \cii\ in nearby Luminous InfraRed Galaxies (LIRGS) also pushes up to around 2\% of the IRAS-based FIR.   We find a spread in \cii/TIR$_{Spitzer}$ for our five galaxies of $0.37\pm0.06\%$ up to $0.86\pm0.20\%$ (using TIR$_{Herschel}$ this becomes $0.73\pm0.11\%$ to $1.19\pm0.31\%$).  

We investigated the possibility of a spatial selection effect as one explanation for the high observed ratios.  The \citetalias{Brauher} comparison sample ratios are averages over the whole galaxy, because the galaxies were unresolved in those observations.  Our new \herschel\ data, on the contrary, can resolve features within the galaxies which may have local enhancements of \cii\ emission. Without complete \cii\ maps of each system we cannot formally disprove that hypothesis, but it seems unlikely given that the observed regions are not unusual in any other band (UV, optical broadband, H$\alpha$, HI, FIR) compared to the rest of the galaxy, and therefore it would be statistically unlikely to pick out locations of enhanced \cii\ by chance in all targets.

Our data follow the slightly decreasing trend of \cii/TIR with increasing $60\mu m/100\mu m$, seen often in the literature (\citealt{Mal97,Luhman98,Leech99,Negishi01}; \citetalias{Brauher}).  The high ratio of \cii -to-TIR for $60\mu m/100\mu m$ values between 0.3-0.6 is discussed briefly by \citetalias{Brauher}.  They propose that either there is a high fraction of intermediate-mass stars that produce copious amounts of UV and \cii\ radiation, or that there is simply a dearth of FIR emission.  The low \oiii/\cii \ ratios observed in our galaxies are consistent with the latter hypothesis, though \cii\ and TIR are both affected by photoelectric heating -- a decrease in one should generally correspond to a decrease in the other.  This suggests that our systems may have a high filling factor of gas with a moderate radiation field G$_0$ and a moderate (diffuse) density, which could lead to elevated levels of \cii\ compared to \oiii\ and TIR.

Our galaxies tend to have similar TIR/B values as found in elliptical galaxies.  However, the high \cii/TIR ratios set them apart.  This combination implies low dust masses in our galaxies.  This is consistent with recent results from \cite{Leroy11}, who demonstrated that dust content in their models decreases with diminishing O/H.

We note that the \cii/TIR ratio seems to follow a weak inverse trend with metallicity, when comparing our sample to other galaxy types.  There is quite a bit of scatter, however, and there is no such trend among irregulars if our sample is omitted.   

\textbf{\oi/TIR}: \textit{Our ratios are consistent with those seen in other samples.}
Again, we see \oi/TIR ratios at the higher end of the \citetalias{Brauher} sample.  There is no strong trend with either FIR color or TIR/B, however.  \citetalias{Mal01} noted that since \oi\ emission is more susceptible to absorption by small interstellar particles, the fact that there is no decrease in \oi/TIR as TIR/B increases could indicate that there is little optical extinction in our sample.  \citetalias{Cormier2015} notes that the DGS sample has a smaller range of \oi/TIR than \cii/TIR, interpreted as possible evidence of \cii\ originating in more than just the PDRs. Our values span roughly the same range, though (both within 0.5\%).

\textbf{\nii/TIR}: \textit{Our values are consistent with the comparison samples.}
\citetalias{Brauher} found all of their data to be below \nii/TIR$\sim$0.001, with some upper limits on the order of 0.006. There were several spiral galaxies listed as detections with ratios down to a few times $10^{-6}$. Our detections fall within this range as well, and overlap well with the spread seen in spiral galaxies.  Our detections broadly agree with the negative trend with $60\mu m/100\mu m$ reported by \citetalias{Mal01}.  The upper limits for DDO 75 and DDO 155, where \nii\ was not detected, suggest that they may lie in the same regime of several extreme spirals from the \citetalias{Brauher} sample.  

There is no appreciable trend with TIR/B among the \textit{detections} in our sample, as their TIR/B values are quite close.  If we were to consider the upper limit for DDO 75, though, a steep negative trend emerges.  There is no obvious trend with metallicity, though there may be a slight increase in \nii/TIR as Z decreases.

\textbf{\oiii/TIR}:  \textit{We see no apparent trends, and our sample is inconsistent with a previously noted relation.}
\citetalias{Brauher} discuss an increase in \oiii/TIR with warmer \iras\ color when considering all galaxy types, attributing this to denser \hii\ regions in warmer galaxies.  Our sample seems to show a slightly \textit{decreasing} trend with 60$\mu$m/100$\mu$m, though the scatter is smaller than found in the \citetalias{Brauher} subsamples.  \citetalias{Cormier2015} notes that the DGS sample has a flattened relation, since \oiii\ can also originate from low density ionized gas.  It seems that there are no overall trends among particular galaxy types with 60$\mu$m/100$\mu$m.  

The \citetalias{Brauher} sample shows a strikingly tight anticorrelation between \oiii/TIR and TIR/B in irregular galaxies, echoed by a similar anticorrelation with L$_{TIR}$ already noted by \citetalias{Cormier2015}. Our galaxies do not follow this -- or any -- trend with TIR/B.  Even if our galaxies were uniformly shifted in TIR/B (which would require a factor of $\sim$150), they would still not line up cleanly with the trend, as the factor of $\sim$3 spread in \oiii/TIR is almost as large as the spread in the \citetalias{Brauher} irregulars.  Our galaxies show a slight decrease in \oiii/TIR as metallicity decreases, however. 

\textbf{(\oi+\cii)/TIR}:  \textit{Our ratios are high relative to other galaxies -- suggesting enhanced photoelectric efficiency in our sample.}
The ratio of (\oi+\cii) to TIR is a diagnostic of gas heating efficiency of PDRs in galaxies, as discussed by \cite{HT97}.  In short, TIR emission traces the gas heating, while \cii\ and \oi\ trace the gas cooling.  Assuming that the photoelectric effect on dust is responsible for the gas heating, the ratio of (\oi+\cii)/TIR is thus a proxy for the photoelectric efficiency.  In the presence of a strong radiation field, dust grains may on average have a slight charge, due to having already lost electrons.  Since it is more difficult to remove additional electrons, the photoelectric efficiency would decrease in those conditions.  Weaker radiation fields, on the other hand, would lead to less ionized dust, and the efficiency would increase. This may explain the high (\oi+\cii)/TIR ratios in our systems.

We note that since \cii\ dominates over \oi, the trend looks almost the same as that in the \cii/TIR comparison. The reference galaxy sample from \citetalias{Brauher}, also similar to that in \cii/TIR, has significant scatter for irregulars and spirals on the lower half of the plot.
\citetalias{Mal01} showed a decreasing trend for this ratio with respect to increasing $60\mu m/100\mu m$, and our data fall nicely in line with this trend, indicating that our sample consists of efficient coolers.  There is no obvious trend with TIR/B within our galaxies, though ignoring the low-end scatter in the \citetalias{Brauher} sample, there may be a slight downward trend overall with other galaxies.

\textbf{I$_{TIR}$ Comparison}: \textit{Values are within expectations, with a possible indication of decreased hydrogen density.} 
Figure~\ref{fig:lineratios_resgrid} in the following section presents spatially resolved measurements of several lines as compared with the total infrared brightness I$_{TIR}$, but for now we only consider the measurements spanning the whole footprint of each galaxy (the hexagon symbols).  At the distances listed in Table~\ref{table:basicinfo}, our TIR brightnesses translate to a span of roughly $10^6 - 10^7$ L$_\sun$, quite small compared to the $10^6 - 10^{12}$ L$_\sun$ range in DGS \citepalias{Cormier2015} and \citetalias{Brauher} samples because we are looking at small regions within faint dwarf galaxies.

Our \cii/TIR and (\oi +\cii)/TIR trends are relatively flat as compared to I$_{TIR}$, though this is consistent with the scatter seen in the DGS sample.  Our five galaxies seem to indicate that \oi/TIR actually increases with I$_{TIR}$ over the observed range for the whole-footprint integrations.  If true, this may indicate that density is dropping - the hydrogen density can decrease by two orders of magnitude from the \oi\ critical density before significantly affecting \cii\ emission.  However, the ranges here are again consistent with the DGS sample scatter.  The apparent increase in \oi/\cii/ with I$_{TIR}$ can be explained by \oi/TIR increasing as \cii/TIR remains roughly flat, and once again is well within the scatter of the DGS sample.

\textbf{Synopsis}:  The \oi/\cii\ ratios in our galaxies are low relative to the \citetalias{Brauher} samples, however the values fall neatly in line with the previously noted trend with dust temperature.   \oiii/\cii\ is generally lower in our sample than for the \citetalias{Brauher} dwarfs, but is consistent with the levels seen in many spiral galaxies.  The \nii/\cii\ values in our dwarfs are extremely low compared to those seen in all the other samples.  The \cii/TIR values in our sample are higher than most other galaxies, while \nii/TIR and \oiii/TIR are normal compared to the other galaxy types.  The one conspicuous outlier considering all of these facts is that \cii\ emission is enhanced relative to both the TIR continuum and to the other lines in our galaxies.  That is, \cii\ could be the primary driver behind many of these findings.  A more detailed comparison of \cii\ and \hi\ will thus be interesting, as will further study of \cii\ relative to star formation tracers, and we will address both of these topics in upcoming papers.

\subsection{Spatially-Resolved Trends}
\label{sec:resolvedscales}

\begin{figure*}
\centering
\includegraphics[width=\textwidth]{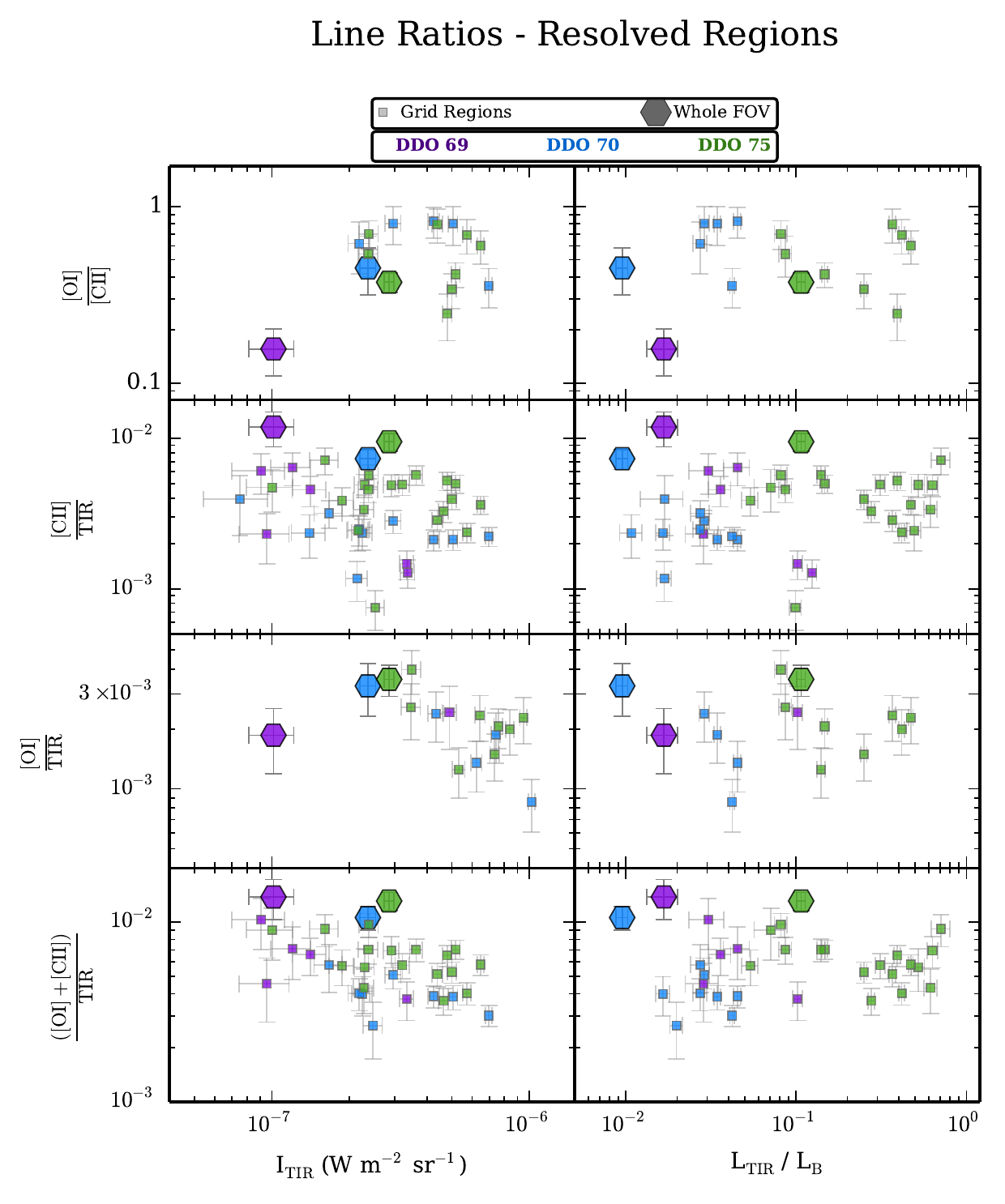}
\caption{\herschel\ line ratio comparison for resolved regions and beam-by-beam grids as described in $\S$~\ref{sec:resolvedscales}. Each resolved data point (square symbols) represents the fluxes contained within a single \cii\ beam width.  Only detections ($3\sigma$) are shown.  We only consider TIR derived from PACS 100$\mu$m maps (DDO 69, DDO 70, DDO 75; squares) for trends on these scales, since these regions are much smaller than the limiting MIPS 160$\mu$m beam size for the \spitzer -based TIR.   The \textit{global} line ratios for TIR calculated from PACS (hexagons) are plotted for reference.     
}
\label{fig:lineratios_resgrid}
\end{figure*}

The spatial arrangement of the line flux peaks bolsters our claim that the bulk of the \cii\ comes from PDRs instead of the diffuse medium in our objects. As discussed in $\S$~\ref{sec:regioncomp}, Figures~\ref{fig:mapcomp} --~\ref{fig:mapcomp3} show that \nii\ emission is usually located at the edge of \cii\ enhancements, while \oi\ peaks are much more closely correlated with carbon.  \cii\ can coexist with both species, but \oi\ originates in PDRs while \nii\ is found only in the more diffuse ionized medium, meaning that the \cii\ observed in our targets may mostly originate in PDRs.  While the spatial argument is not a proof on its own, it does complement the evidence from the \nii/\cii\ ratio.

The various FIR line ratios in discrete resolved regions within our galaxies are compared to TIR/B as well as integrated TIR brightness in Figure~\ref{fig:lineratios_resgrid}.  As discussed in \S~\ref{sec:TIRvsFIR}, we make a grid of integration regions by dividing the maps into uniform divisions the size of the \cii\ beam.  The MIPS 160$\mu$m data, which limit the resolution of the TIR luminosity, have a PSF that is nearly the size of each \herschel\ footprint, meaning several of the individually resolved and gridded regions would be contained within one \spitzer\ TIR beam. For this reason, only \herschel-based TIR values are considered for these smaller regions.   Since the \oiii\ and \nii\ maps require spatial integration over specific areas to recover detections above 3$\sigma$ significance, there are no regions where both the line flux and TIR flux within one of the 5$\times$5 beam-sized apertures have S/N$>$3 for these particular lines.  For this reason, we omit them from the plots in Figure ~\ref{fig:lineratios_resgrid}.

Generally speaking, there are no obvious trends for any of the resolved line ratios across the data of all three galaxies in concert.  There are a few correlations for ratios among the data within the footprints of individual galaxies, though.

\oi/\cii\ does not vary appreciably as either the TIR brightness or TIR/B change within our footprints, though there is a slight decrease within each galaxy, if anything.  As all the line ratio values are less than one, \cii\ dominates cooling over \oi\ in all detected areas. For both DDO 70 and DDO 75, there are a few errant regions where \oi/\cii\ drops to $\sim$0.3 from the average $\sim$0.7, but not in a predictive manner for larger TIR values.  Relatively stable ratios of these lines are more or less expected, since the spatial comparison maps (Figures ~\ref{fig:mapcomp} --~\ref{fig:mapcomp3}) show qualitatively that their emission tends to agree well.  Changes in the \oi/\cii\ ratio indicate variation in density and radiation field  -- \cii\ has a higher ionization potential, but lower critical density, than \oi .  So a slight decrease in \oi/\cii\ could indicate that some of the regions with both species of emission are slightly less dense (decrease in \oi) or have slightly softer radiation (larger \cii\ fraction).  

The \cii/TIR and (\oi + \cii)/TIR plots look similar, since \cii\ dominates the cooling over \oi\ in our galaxies.   DDO 75 shows little sign of any change in the ratios as either TIR brightness or TIR/B vary within our maps.  DDO 69, however, shows definite decreases with both I$_{TIR}$ and TIR/B.  The DDO 70 ratios decrease slightly with TIR brightness, to a lesser extent than DDO 69, but do not change much with TIR/B.

\oi/TIR decreases with I$_{TIR}$ for DDO 70 and DDO 75, however the dependence is not strong - omitting the single highest \oi/TIR value from DDO 75 leaves the trend less tenable. There is again an apparent decreasing tendency among the resolved \oi/TIR measurements as TIR/B increases for those two galaxies, but this is weak considering the scatter and error bars.  The dispersion among \oi/TIR ratios is smaller than for \cii/TIR by about half, though this could simply be a bias since there are fewer \oi\ detections than \cii .

\section{SUMMARY}
\label{sec:conclusion}

We have presented \herschel\ PACS spectroscopy of a sample of five dwarf galaxies from the \lt\ Survey --  DDO 69, DDO 70, DDO 75, DDO 155, and WLM.  All five galaxies are faint, low-metallicity (7.4 $<$ Z $<$ 7.8), nearby (between 0.8 and 2.2 Mpc) dwarfs with moderate star formation (averaging 0.09 M$_\odot$ yr$^{-1}$).  Prior to \herschel, these types of systems were extremely difficult to observe in the FIR, but these recent detections now make it possible to explore the regime of dwarfs with very low metallicity and not undergoing starburst episodes, on the scales of large star-forming complexes.  

We observed four fine-structure lines in each galaxy: \cii 158$\mu$m, \oi 63$\mu$m, \oiii 88$\mu$m, and \nii 122$\mu$m, with the exception of DDO 75 which had no \oiii\ observations.  \cii\ and \oi\ are important species in the PDR cooling process, while \oiii\ and \nii\ are indicators of regions with ionized gas.  We compare to the total infrared continuum as estimated by \spitzer\ 24, 70, and 160$\mu$m maps, though the coarse resolution of the MIPS maps effectively limits TIR$_{\spitzer}$  to average values over the field of view of the PACS maps.  Where available, we also utilize new PACS 100$\mu$m observations to estimate TIR at the same resolution as the line emission.  

The dwarfs in our sample have low TIR/B as well as low $60\mu m / 100\mu m$.  This indicates cooler dust in lower quantities on average in our systems than for other dIrrs and spirals.  Thus they are at one extreme, having low temperature, low dust content, and low metallicity (5-13\% of Z$_\sun$). 

Upon examination of several line ratios and comparisons with previous studies of other galaxies, we find:

\begin{itemize}

\item The \cii/TIR ratios in all our galaxies are extremely high, on the order of 0.4-1\%. This is comparable to some of the highest ratios detected in other galaxies, including dwarfs, spirals, ellipticals, and LIRGS.  \oiii/TIR and \nii/TIR are relatively normal when compared to other galaxy types.

\item Our galaxies have low observed \oi/\cii\ ratios, smaller than most other dIrrs, but still following a previously noted trend with FIR color. This, combined with low $60\mu m / 100\mu m$ and high (\oi+\cii)/TIR values, suggests cool dust temperatures and efficient cooling of gas across our systems.

\item \oiii/\cii\ is low compared to many other dwarfs, and is consistent with the upper range of spirals in the \citetalias{Brauher} sample. Our \oiii/\oi\ ratios also span an intermediate range between the clustered irregular and spiral values.  

\item  \nii\ emission, which comes from ionized regions, is generally not widely detected in our observations.  Furthermore, detected regions are typically located to the side of areas bright in \cii\ rather than directly overlapping.   This implies that the \cii\ emission is primarily coming from PDRs rather than \hii\ regions, or that N/C abundances are not as expected, or both.  

\item Our systems may have softer radiation fields than other dwarfs, suggested by low limits to \oiii/\nii\ ratios when compared to the DGS galaxies.

\item We are able to resolve FIR fine structure lines on scales of 45-123 pc, or the size of large star-forming complexes.  Due to the extreme faintness of our targets, only the \cii\ and \oi\ lines have enough signal to be reliably detected on this scale. \cii/TIR, \oi/TIR and (\oi +\cii)/TIR slightly decrease with increasing TIR brightness within each galaxy (most noticeably for DDO 69).  

\item \oi\ is generally widespread and coincident with the \cii\ in our maps.  \oiii\ peaks where \hii\ regions are located, as expected.  The \cii, rather than explicitly mirroring the star formation tracers, tends to follow the \hi\ emission more closely, which may allow us to proble the CO-dark H$_2$ in future work.

\end{itemize}

These resolved line ratios will help us to model the properties of extragalactic PDRs in low metallicity systems at high spatial resolution in future publications.

\acknowledgments{{\it Acknowledgments:}  Funding for this project was provided by NASA JPL RSA grant 1433776.  SCM acknowledges support from the Agence Nationale de la Recherche (ANR) through the programme SYMPATICO (Program Blanc Projet ANR-11-BS56-0023).  VH acknowledges support from the Science and Technology Facilities Council (STFC) under grant ST/J001600/1.  \herschel\ is an ESA space observatory with science instruments provided by European-led Principal Investigator consortia and with important participation from NASA.  This research has made use of the NASA/IPAC Extragalactic Database (NED), which is operated by the Jet Propulsion Laboratory, California Institute of Technology, under contract with the National Aeronautics and Space Administration.

Facilities: \facility{\herschel\ (PACS)}, \facility{Very Large Array}

\bibliographystyle{aj}


\end{document}